%% file: szpol_hiell.tex
\documentclass[twocolumn]{aastex62}
\usepackage{amsmath}
\usepackage{natbib}
\usepackage{graphicx}
\usepackage{epsf}
\usepackage{color}
\input{epsf}
\usepackage{epsfig}
\usepackage{rotating}
\usepackage{afterpage}
\DeclareGraphicsExtensions{.jpg,.pdf,.png,.eps,.ps}
\graphicspath{{FIGURES/},{.}}
\usepackage{lipsum}
\usepackage{multirow}

\newcommand{\lcdm}{\mbox{$\Lambda$}CDM}
\newcommand{\planck}{{\it Planck}}
\newcommand{\wmap}{{\it WMAP}}

\newcommand{\sptpol}{SPTpol}

\newcommand{\sqdeg}{\mbox{deg$^2$}}
\newcommand{\chisq}{\ensuremath{\chi^2}}
\newcommand{\delchisq}{\ensuremath{\Delta\chi^2}}

\newcommand{\ltsima}{$\; \buildrel < \over \sim \;$}
\newcommand{\ltsim}{\lower.5ex\hbox{\ltsima}}
\newcommand{\tbd}{{\bf \textcolor{red}{TBD}}}

\newcommand{\sptsz}{SPT-SZ}

\newcommand{\ukarcmin}{\ensuremath{\mu{\rm K-arcmin}}}

\newcommand{\uksq}{\ensuremath{\mu {\rm K}^2}}
\newcommand{\dksz}{\ensuremath{D_{3000}^{\rm kSZ}}}
\newcommand{\dpksz}{\ensuremath{D_{3000}^{\rm p-kSZ}}}
\newcommand{\dhksz}{\ensuremath{D_{3000}^{\rm h-kSZ}}}
\newcommand{\dtsz}{\ensuremath{D_{3000}^{\rm tSZ}}}

\newcommand\comment[1]{}

\include{mcmc_results}
\include{r19_baseline_parameters}
\include{r19_altsz_parameters}

\include{r19_ksz_eor_parameters}
\include{r19_probabilities}

\newcommand{\degsq}{deg$^2$}

\begin{document}

\title{An Improved Measurement of the Secondary Cosmic Microwave Background Anisotropies from the \sptsz{} $+$ \sptpol{} Surveys}

\input{authors}

\email{christian.reichardt@unimelb.edu.au}
 
\begin{abstract}
We report new measurements of millimeter-wave power spectra in the angular multipole range $2000 \le \ell \le 11,000$ (angular scales $5^\prime \gtrsim \theta \gtrsim 1^\prime$).  
By adding 95 and 150\,GHz data from the low-noise 500\,\sqdeg{} \sptpol{} survey to the \sptsz{} three-frequency 2540\,\sqdeg{} survey, we substantially
reduce the uncertainties in these bands. 
These power spectra include contributions from the primary cosmic microwave background, cosmic infrared background, radio galaxies, and thermal and kinematic Sunyaev-Zel'dovich (SZ) effects. 
The data favor a thermal SZ (tSZ) power at 143\,GHz of $D^{\rm tSZ}_{3000} = \DtSZ{}\,\uksq{}$ and a kinematic SZ (kSZ) power of $D^{\rm kSZ}_{3000} = \DkSZ{}\, \uksq{}$. 
This is the first measurement of kSZ power at $\ge3\,\sigma$. 
We study the implications of the measured kSZ power for the epoch of reionization, finding the duration of reionization to be $\Delta z_{re} = \DzEorCalabreseFlatBispecInterval$ ($\Delta z_{re}<\DzEorCalabreseFlatBispec$ at 95\% confidence), when combined with our previously published tSZ bispectrum measurement.
\end{abstract}


\bigskip\bigskip

\section{Introduction}
\setcounter{footnote}{0}

The cosmic microwave background (CMB) is best known for providing a snapshot of the early Universe. 
However, on small angular scales, secondary anisotropies in the CMB, created by interactions between CMB photons and large-scale structure, also provide clues about the late-time Universe. 
In particular, these secondary anisotropies encode information about the amplitude of structure growth and duration of the epoch of reionization.

The most significant secondary anisotropies at angular scales of a few arcminutes are the kinematic and thermal Sunyaev-Zel'dovich (SZ) effects. 
Both SZ effects are due to CMB photons scattering off of free electrons along their path. 
The kinematic SZ (kSZ) effect is due to an induced doppler shift in the scattered photons, and thus the kSZ signal from a given volume element is proportional to $(v/c) n_e$ where $v$ is the bulk velocity of the electrons, and $n_e$ is the number density of free electrons. 
The kSZ power spectrum is expected to have significant contributions from the epoch of reionization due to the large contrasts in ionization fraction as the Universe reionizes  \citep{gruzinov98,knox98}, and  at late times when there are larger relative velocities and density contrasts \citep[e.g.,][]{shaw12,battaglia13c}.

In contrast, the thermal SZ (tSZ) effect is due to the energy transfer from hot electrons to the colder CMB photons, and has a signal amplitude of $(k_B T_e/m_ec^2)n_e$, where $m_e$ is the mass of the electron and $T_e$ is the temperature of the electrons.  
While the kSZ effect does not change the CMB spectrum, the net energy transfer to the photons in the tSZ effect translates to a reduction in the number of CMB photons below 217\,GHz as these photons are up-scattered towards higher frequencies. 
One can use the difference in how the tSZ and kSZ effects scale with frequency to simultaneously measure both terms.  
The tSZ anisotropy signal scales steeply with the normalization of the matter power spectrum, which can be
parametrized by $\sigma_8$, the RMS of the $z=0$ linear mass distribution on 8$h^{-1}$ Mpc scales,  \citep[e.g.,][]{komatsu02}.

The secondary CMB anisotropies are not the only sources of anisotropy in millimeter-wave maps on arcminute scales. 
Galaxies also emit at these wavelengths, both synchrotron-dominated active galactic nuclei (AGN, e.g., \citealt{dezotti10}) and thermal dust emission from dusty, star-forming galaxies (DSFGs, e.g., \citealt{planck11-6.6, mocanu13}, Everett, et al., in prep.).
While the brightest of these sources can be individually detected and masked, it is impossible to remove all of the fainter galaxies as there are many such DSFGs within each square arcminute \citep{lagache05,casey14}. 
The DSFG signal can be split between a term that does not spatially cluster (the ``Poisson'' component) plus a spatially clustered term \citep{viero13a}
We can separate the AGN and DSFGs from the SZ effects using both angular and spectral information. 

\subsection{Previous measurements}
\label{sec:intro_prevmeas}

Measurements of the millimeter sky at arcminute scales have been made by both the Atacama Cosmology Telescope (ACT; \citealt{das11b, das14}) and South Pole Telescope (SPT) SZ survey \citep{lueker10, shirokoff11, reichardt12b, george15}. 
The ACT collaboration \citep{das14,dunkley13} measured  $D^{\rm tSZ}_{3000} = 3.3\pm1.4\,\mu{\rm K}^2$ and $D^{\rm kSZ}_{3000} < 8.6\,\mu{\rm K}^2$ (95\% CL) at 150\,GHz and $\ell=3000$. 
The final SPT-SZ bandpowers reported by \citet[][hereafter G15]{george15} led to even tighter constraints on the tSZ power at 143\, GHz of $D^{\rm tSZ}_{3000} = 4.08^{+0.58}_{-0.67}\,\uksq{}$ and on the kSZ power of $D^{\rm kSZ}_{3000} = 2.9 \pm 1.3\, \uksq{}$. 
On larger scales, $\ell \le 2000$, the \planck{} collaboration made a high-significance detection of the tSZ power spectrum \citep{planck13-21, planck15-22}.
The observed tSZ power is consistent across all three experiments.

The data used to constrain the tSZ and kSZ power spectra can also teach us about the cosmic infrared background (CIB), radio galaxies and correlation between the CIB and galaxy clusters. 
G15 detected a non-zero correlation between the CIB and galaxy clusters, modelled as a constant, at a significance of more than $3\,\sigma$, finding $\xi = 0.113^{+0.057}_{-0.054}$.

\subsection{This work}
This work adds data from the low-noise 500\,\sqdeg{} \sptpol{} survey to the 2540\,\sqdeg{} SPT-SZ survey maps used by G15. 
The \sptpol{} data substantially reduces the map noise at 95 and 150\,GHz over the 500\,\sqdeg{} that was observed by both surveys, however the 220\,GHz maps are unchanged from G15 since \sptpol{} did not observe at 220\,GHz. 
The lower noise levels at 95\,GHz yield a three-fold reduction in the bandpower uncertainties at 95$\times$95\,GHz; the improvement is more modest ($\sim$30\%) but still significant at 150$\times$150\,GHz.

The outline of this work is as follows. 
We review the observations and power spectrum analysis in \S\ref{sec:dataanalysis}. 
Systematics checks done on the data are described in \S\ref{sec:jackknife}, before the bandpowers are presented in \S\ref{sec:bandpowers}.
We discuss the modelling of the bandpowers in \S\ref{sec:model}, and the constraints on this model in \S\ref{sec:results}. 
We explore the implications for the epoch of reionization in \S\ref{sec:kszinterp} before concluding in  \S\ref{sec:conclusions}.

\section{Data and analysis}
\label{sec:dataanalysis}

We present power spectra from the combined \sptsz{} and \sptpol{} surveys at 95, 150, and 220\,GHz. 
We use a pseudo-${\it C}_\ell$ cross-spectrum method \citep{hivon02,polenta05,tristram05} to estimate the power spectra. 
The data is calibrated by comparing to the \emph{Planck} 2015 CMB maps.

\subsection{Data}
\label{subsec:data}

This work uses data from the \sptsz{} and \sptpol{} cameras on the South Pole Telescope. 
Details on the telescope and cameras can be found in  \citet{ruhl04}, \citet{padin08}, \citet{shirokoff09}, \citet{carlstrom11}, \citet{henning12}, \citet{sayre12}, and \citet{austermann12}.

As described by G15, the 2540 \degsq\ \sptsz{} survey was conducted from 2008 to 2011. 
The survey region was split into 19 contiguous sub-patches, referred to as fields, for observations.  
The specific field locations and extents can be found in Table 1 of \citet{story13}, hereafter S13. 
The \sptpol{} 500\,\sqdeg{} survey fully or partially overlaps six of these 19 fields.  
Bandpowers for the 13 non-overlapping fields are identical to G15 (except for an updated calibration, see \S\ref{sec:beamcal}). 

We treat the overlapping region as a single field, and coadd the time-ordered data (TOD) from both \sptpol{} and \sptsz{} data into maps. 
Details of the time-ordered data (TOD), filtering, and map-making can be found in \citet{shirokoff11} for the \sptsz{} data and in \citet{henning18} for the \sptpol{} data. 
The \sptpol{} filtering options have been tuned to closely match the \sptsz{} maps used by G15. 
After combining data from the full 2540\,\degsq{}, the approximate statistical weight from the new \sptpol{} data is 83\% at 95\,GHz, 44\% at 150\,GHz, and 0\% at 220\,GHz.

\subsection{Beams and calibration}
\label{sec:beamcal}

The \sptsz{} beams are measured using a combination of bright point sources in each field, Venus, and Jupiter as described in \citet{shirokoff11}. 
The \sptpol{} beams are measured using Venus alone as described by \citet{henning18}. 
We take a weighted average, based on the statistical weight of each dataset in the map, of the beams from the two experiments to estimate the effective beam of the combined survey. 
Note that the final bandpowers should be robust to an error in this effective beam calculation since the transfer function simulations (\S\ref{sec:sims}) use the correct beams for each period of data. 
For both experiments, the main lobes of the beam  are well-represented by 1.7$^\prime$, 1.2$^\prime$, and 1.0$^\prime$ FWHM Gaussians at 95, 150, and 220$\,$GHz respectively.

We use the absolute calibration factors calculated by \citet{hou18} and \citet{mocanu19} for the \sptsz{} data and the absolute calibration from \citet{henning18} for the \sptpol{} data. 
In both cases, the calibration is determined by comparing the \sptsz{} (or \sptpol) maps with \planck{} maps in the same region of sky. 
The uncertainties are correlated between frequency bands due to sample variance. 
The final uncertainties in power are  [0.33\%, 0.18\%, 0.42\%] at  [95, 150,  220] GHz.

The treatment of the beam and calibration uncertainties in the parameter estimation is described in \S\ref{sec:beamunc}.

\subsection{Power spectrum estimation}
\label{sec:bandpowerest} 

Following G15, we use a pseudo-$C_\ell$ method to estimate the power spectrum \citep{hivon02}.  
 Pseudo-$C_\ell$ methods start by calculating a (biased) power spectrum from the Fourier transform of the map (in flat-sky), and then correct this biased spectrum for effects such as TOD filtering, beams, and finite sky coverage  \citep{hivon02}. 
Following \citet{polenta05, tristram05}, we use cross-spectra instead of auto-spectra to avoid noise bias in the result. 
 We report the power spectrum in terms of $\mathcal{D}_\ell$, where
\begin{equation}
\mathcal{D}_\ell=\frac{\ell\left(\ell+1\right)}{2\pi} C_\ell\;.
\end{equation}

More details on the power spectrum estimator can be found in previous \sptsz{} papers: e.g.,  \citet{lueker10}, \citet[][hereafter R12]{reichardt12b}, and G15. 
We emphasize that for the 13 non-overlapping fields, this work simply reuses the G15 bandpowers for each field. 
We briefly describe the method in the following sections, focusing on the part that is new in this work -- the power spectrum estimation for the combined \sptsz{} + \sptpol{} maps.

\subsubsection{Cross spectra}
\label{sec:xspec}

Before Fourier transforming the maps, we apply a window to each map that smoothly goes to zero at the map edges. 
The window also masks point sources above $6.4$\,mJy at 150\,GHz from the source catalog in Everett et al., in prep. 
The mask for each point source has a 2 arcmin radius disc for sources detected with $S_{\rm 150\,GHz} \in [6.4, 50]$\,mJy, and  a 5 arcmin radius disk for sources above 50\,mJy. 
In both cases,  a Gaussian taper with $\sigma_{taper}=5$ arcmin is applied outside the radius of the disk. 
For the combined \sptpol{} and \sptsz{} field which has anisotropic noise due to variations in the amount of integration time, this window also preferentially weights  the lower noise regions. 

After Fourier transforming the windowed maps, we take the weighted average of the two-dimensional power
spectrum within an $\ell$-bin $b$,
\begin{equation}
\label{eqn:ddef}
 \widehat{D}^{\nu_i \times \nu_j, AB}_b\equiv \left< \frac{\ell(\ell+1)}{2\pi}\mathrm{Re}\left[\tilde{m}^{\nu_i,A}_{\ell}\tilde{m}^{\nu_j,B*}_{\ell}\right] \right>_{\ell \in b}, 
\end{equation} 
where $\tilde{m}^{\nu_i, A}$ is the Fourier transformed map. 
Here, $A, B$ are the observation indices, while $\nu_i, \nu_j$ are the observation frequencies (e.g., 150\,GHz). 
We average all cross-spectra $\widehat{D}^{AB}_b$ that have $A \neq B$ to get the  binned power spectrum $\widehat{D}_b$.  
As in R12, we eliminate the noisier modes along the scan direction by excluding modes with $\ell_x < 1200$. 
We refer to the binned power, $\widehat{D}_b$, as a ``bandpower."

\subsubsection{Simulations}
\label{sec:sims}

The transfer function as well as sample variance for the combined \sptpol{} and \sptsz{} field is calculated from a suite of 200 signal-only simulations. 
We convolve the simulated skies by the measured beam for each frequency and observing year before sampling the realizations based on the pointing information. 
The simulated TOD are filtered and binned into maps in the same way as the real data.  

The simulated skies include  Gaussian realizations of the best-fit lensed \planck{} 2013 $\Lambda$CDM primary CMB model, SZ models and extragalactic source contributions. 
Following G15, the kSZ power spectrum is based on the \citet{sehgal10} simulations with an amplitude of $2.0\, \mu{\rm K}^2$ at $\ell=3000$. 
The tSZ power spectrum is taken from the \citet{shaw10} simulations, normalized to have an amplitude of $4.4\, \mu{\rm K}^2$ at $\ell=3000$ at 153\,GHz. 
The extragalactic source term can be split into three components: spatially clustered and Poisson-distributed DSFGs, and Poisson-distributed radio galaxies. 
Motivated by the predictions of the \citet{dezotti05} model for a 6.4 mJy flux cut at 150\,GHz, the radio power is set to $D_{3000}^{r} = 1.28\, \mu{\rm K}^2$ at $150\,$GHz. 
We assume a radio spectral index of $\alpha_r=-0.53$\footnote{i.e. the radio source flux in Jy is proportional to $\nu^{\alpha_r}$, where $\nu$ is the frequency.} and 1-sigma scatter on the spectral index of 0.1. 
 The DSFG Poisson power is set to $7.54\,\mu{\rm K}^2$ at 154\,GHz with a modified black-body spectrum\footnote{i.e. the dusty galaxy flux in Jy is proportional to $\nu^{\beta} B_\nu(T_{\rm dust})$} with T$_{\rm dust}$=12\,K and $\beta=2$.
 The clustered DSFG component is modeled by a $D_\ell \propto \ell^{0.8}$ term normalized to $D_{3000}^{c} = 6.25 \,\mu{\rm K}^2$ and the same spectral dependence as the Poisson DSFG. 
 These simulations do not include non-Gaussianity in the tSZ, kSZ, and radio source signals and therefore slightly underestimate the sample variance. 
\citet{millea12} argue the non-Gaussian sample variance of the SZ and radio terms is negligible since the instrumental noise power is always larger than these terms.

\subsubsection{Covariance estimation and conditioning}
\label{sec:cov}

In order to compare the measured bandpowers to theory, we need to estimate a covariance matrix including both sample variance and instrumental noise variance.  
As in R12 and G15, the sample variance is estimated from signal-only  simulations (\S\ref{sec:sims}),  and the noise variance is empirically determined from the distribution of the cross-spectrum bandpowers $D^{\nu_i\times\nu_j,AB}_b$ between observations A and B, and frequencies $\nu_i$ and $\nu_j$. 
A noisy estimate of the bandpower covariance matrix could degrade parameter constraints \citep[see, e.g.,][]{dodelson13}. 
Thus we  follow G15 and ``condition'' the covariance matrix to minimize the noise on the covariance estimate and largely avoid this degradation.

The covariance matrix depends on the signal power, and, if both bandpowers share a common map, noise power. 
As the errors on the off-diagonal elements include terms proportional to the (potentially much larger) diagonal elements, the uncertainty on the off-diagonal elements can be large compared to the true covariance. 
As a result, we estimate these values analytically from the diagonal elements using the equations in Appendix A of L10. 

\subsubsection{Field weighting}
\label{sec:fieldweighting}

We follow G15 and  weight  each field and frequency cross spectrum  based on the average of the inverse of the diagonal of the covariance matrix over the bins $2500<\ell<3500$. 
These weights adjust for the differences in noise and sample variance between fields; beam and calibration errors are deliberately not included. 
As argued by G15, the angular range, $2500<\ell<3500$, is where the data have the most sensitivity to SZ signals.

We calculate the combined bandpowers, $D_b$, as:
\begin{equation}
D_b = \sum_{i}D_{b}^{i}w^{i}, 
\end{equation}
where $D_{b}^{i}$ is the bandpower of field $i$ and $w_i$ the weight. 
The covariance matrix likewise can be expressed:
\begin{equation}
\textbf{C}_{bb^\prime} = \sum_{i}w^{i}\textbf{C}_{bb^\prime}^{i}w^{i}.
\end{equation}
The sum of the weights is normalized to unity.

\subsubsection{Beam and calibration uncertainties}
\label{sec:beamunc}

To handle the calibration uncertainties, we include three calibration factors in the parameter fitting, one per frequency. 
We marginalize over these three factors, with a prior based on the measured calibration uncertainty for all parameter fits. 

We follow \citet{aylor17} for the treatment of beam uncertainties. 
The beam correlation matrix, $\rho^{\rm beam}_{bb^\prime}$ is calculated as described by G15, using the fractional beam errors for each year and the relative weights of each year of data over the SPT-SZ and SPTpol surveys. 
At each step in the chain, we use the predicted theory bandpowers ($D^{\rm theory}_{b}$) to convert this beam correlation matrix into a beam covariance according to: 
\begin{equation}
\textbf{C}^{\rm beam}_{bb^\prime} = \pmb{\rho}^{\rm beam}_{bb^\prime}D^{\rm theory}_{b}D^{\rm theory}_{b^\prime}.
\end{equation}
We  add this beam covariance to the bandpower covariance matrix which contains the effects of sample variance, and instrumental noise. 
The likelihood for that specific theoretical model is then evaluated using this combined covariance matrix. 

\begin{table*}[ht!]
\begin{center}
\caption{\label{tab:bandpowers} Bandpowers}
\small
\begin{tabular}{cc|rr|rr|rr}
\hline\hline
\rule[-2mm]{0mm}{6mm}
& &\multicolumn{2}{c}{$95\,$GHz} &\multicolumn{2}{c}{$150\,$GHz} & \multicolumn{2}{c}{$220\,$GHz} \\
$\ell$ range&$\ell_{\rm eff}$ &$\hat{D}$ ($\mu{\rm K}^2$)& $\sigma$ ($\mu{\rm K}^2$) &$\hat{D}$ ($\mu{\rm K}^2$)& $\sigma$ ($\mu{\rm K}^2$)&$\hat{D}$ ($\mu{\rm K}^2$)& $\sigma$ ($\mu{\rm K}^2$) \\
\hline

2001 -  2200 &  2077   & 218.4 &    3.8   & 215.6 &    2.3   & 286.2 &    6.5   \\ 
2201 -  2500 &  2332   & 128.2 &    1.9   & 125.9 &    1.1   & 201.7 &    4.3   \\ 
2501 -  2800 &  2636   &  81.9 &    1.1   &  80.29 &   0.67   & 170.4 &    4.1   \\ 
2801 -  3100 &  2940   &  52.84 &   0.79   &  51.88 &   0.46   & 156.9 &    4.0   \\ 
3101 -  3500 &  3293   &  36.87 &   0.58   &  36.89 &   0.31   & 155.4 &    3.7   \\ 
3501 -  3900 &  3696   &  31.35 &   0.57   &  31.19 &   0.29   & 182.8 &    4.4   \\ 
3901 -  4400 &  4148   &  28.85 &   0.65   &  31.24 &   0.29   & 202.0 &    4.8   \\ 
4401 -  4900 &  4651   &  30.25 &   0.89   &  33.62 &   0.35   & 245.7 &    6.0   \\ 
4901 -  5500 &  5203   &  35.3 &    1.1   &  39.73 &   0.42   & 290.2 &    7.0   \\ 
5501 -  6200 &  5855   &  43.4 &    1.9   &  46.34 &   0.53   & 349.8 &    8.7   \\ 
6201 -  7000 &  6607   &  44.6 &    3.2   &  57.24 &   0.72   &   435. &    11.   \\ 
7001 -  7800 &  7408   &  46.7 &    6.7   &  69.5 &    1.2   &   524. &    15.   \\ 
7801 -  8800 &  8310   &    61. &    12.   &  89.0 &    1.8   &   665. &    21.   \\ 
8801 -  9800 &  9311   & - & -   &  98.7 &    2.9   &   729. &    34.   \\ 
9801 - 11000 & 10413   & - & -   & 122.0 &    4.5   &   962. &    49.   \\ 
\comment{ 
 2001 -  2200 &  2106 & 212.4 &   6.3 & 207.1 &   6.3 &  271.6 &  23.4 \\ 
 2201 -  2500 &  2357 & 128.3 &   4.0 & 121.3 &   3.8 &  190.3 &  16.7 \\ 
 2501 -  2800 &  2657 &  80.9 &   2.9 &  77.6 &   2.6 &  161.0 &  14.5 \\ 
 2801 -  3100 &  2958 &  52.8 &   2.4 &  50.4 &   1.8 &  151.5 &  14.3 \\ 
 3101 -  3500 &  3308 &  40.5 &   2.2 &  36.2 &   1.4 &  151.0 &  14.4 \\ 
 3501 -  3900 &  3709 &  32.0 &   2.5 &  30.8 &   1.2 &  174.9 &  17.1 \\ 
 3901 -  4400 &  4159 &  33.5 &   3.0 &  31.2 &   1.3 &  198.8 &  19.5 \\ 
 4401 -  4900 &  4660 &  33.0 &   4.2 &  33.7 &   1.4 &  239.9 &  23.3 \\ 
 4901 -  5500 &  5210 &  35.5 &   5.8 &  40.5 &   1.7 &  274.9 &  26.2 \\ 
 5501 -  6200 &  5861 &  50.7 &   8.8 &  47.2 &   1.9 &  339.1 &  32.0 \\ 
 6201 -  7000 &  6612 &  30.2 &  15.8 &  58.3 &   2.3 &  419.8 &  39.6 \\ 
 7001 -  7800 &  7412 &  52.0 &  29.1 &  72.3 &   3.2 &  488.0 &  47.2 \\ 
 7801 -  8800 &  8313 & 109.7 &  52.6 &  93.9 &   4.5 &  638.1 &  63.2 \\ 
 8801 -  9800 &  9313 &   - &   - &  97.6 &   6.2 &  710.5 &  76.5 \\ 
 9801 - 11000 & 10413 &   - &   - & 123.2 &   9.4 & 1010.4 & 111.1 \\ }

\hline
&&\multicolumn{6}{c}{}\\
& & \multicolumn{2}{c}{$95\times150\,$GHz} & \multicolumn{2}{c}{$95\times220\,$GHz} & \multicolumn{2}{c}{$150\times220\,$GHz} \\
\hline

2001 -  2200 &  2077   & 213.3 &    2.9   & 207.2 &    4.0   & 225.9 &    2.9   \\ 
2201 -  2500 &  2332   & 123.5 &    1.4   & 121.6 &    2.2   & 140.7 &    1.6   \\ 
2501 -  2800 &  2636   &  76.72 &   0.82   &  77.7 &    1.6   &  98.8 &    1.2   \\ 
2801 -  3100 &  2940   &  47.73 &   0.54   &  50.0 &    1.4   &  73.03 &   1.00   \\ 
3101 -  3500 &  3293   &  32.01 &   0.36   &  34.2 &    1.2   &  61.78 &   0.80   \\ 
3501 -  3900 &  3696   &  24.38 &   0.34   &  26.6 &    1.4   &  63.77 &   0.87   \\ 
3901 -  4400 &  4148   &  22.47 &   0.35   &  28.3 &    1.5   &  70.62 &   0.89   \\ 
4401 -  4900 &  4651   &  22.46 &   0.46   &  32.0 &    2.0   &  82.4 &    1.1   \\ 
4901 -  5500 &  5203   &  25.00 &   0.58   &  35.1 &    2.6   &  97.1 &    1.3   \\ 
5501 -  6200 &  5855   &  28.88 &   0.79   &  47.1 &    3.2   & 116.6 &    1.6   \\ 
6201 -  7000 &  6607   &  34.9 &    1.2   &  55.8 &    5.2   & 149.8 &    2.1   \\ 
7001 -  7800 &  7408   &  39.2 &    2.0   &  60.4 &    8.7   & 179.1 &    3.1   \\ 
7801 -  8800 &  8310   &  45.8 &    3.3   &    75. &    13.   & 224.0 &    4.3   \\ 
8801 -  9800 &  9311   &  74.8 &    6.5   &    87. &    29.   & 276.1 &    7.2   \\ 
9801 - 11000 & 10413   &    83. &    14.   &   203. &    62.   &   351. &    11.   \\ 
\comment{ 
 2001 -  2200 &  2106 & 205.8 &   5.6 & 200.3 &   9.6 &  216.0 &  11.2 \\ 
 2201 -  2500 &  2357 & 120.8 &   3.4 & 118.3 &   5.8 &  134.9 &   7.2 \\ 
 2501 -  2800 &  2657 &  74.7 &   2.3 &  75.9 &   4.0 &   95.2 &   5.2 \\ 
 2801 -  3100 &  2958 &  46.9 &   1.6 &  50.1 &   3.1 &   70.4 &   4.1 \\ 
 3101 -  3500 &  3308 &  32.3 &   1.2 &  33.8 &   2.5 &   60.1 &   3.6 \\ 
 3501 -  3900 &  3709 &  24.0 &   1.0 &  27.2 &   2.7 &   61.7 &   3.8 \\ 
 3901 -  4400 &  4159 &  22.8 &   1.1 &  26.0 &   3.0 &   68.6 &   4.3 \\ 
 4401 -  4900 &  4660 &  22.3 &   1.3 &  32.5 &   3.9 &   79.2 &   5.0 \\ 
 4901 -  5500 &  5210 &  26.3 &   1.5 &  34.5 &   4.6 &   96.8 &   5.9 \\ 
 5501 -  6200 &  5861 &  29.1 &   2.0 &  49.6 &   6.4 &  113.5 &   6.8 \\ 
 6201 -  7000 &  6612 &  34.6 &   2.9 &  54.5 &   9.2 &  151.2 &   8.9 \\ 
 7001 -  7800 &  7412 &  39.9 &   4.8 &  66.2 &  14.8 &  176.6 &  10.7 \\ 
 7801 -  8800 &  8313 &  43.3 &   7.9 &  80.8 &  23.3 &  224.4 &  14.1 \\ 
 8801 -  9800 &  9313 &  92.5 &  16.0 & 124.8 &  44.6 &  280.0 &  19.2 \\ 
 9801 - 11000 & 10413 &  85.8 &  31.8 & 123.9 &  83.0 &  352.4 &  26.0 \\
 }

\hline
\end{tabular}
\tablecomments{Angular multipole range, weighted multipole value $\ell_{\rm eff}$, bandpower $\hat{D}$, 
and bandpower uncertainty $\sigma$ for the six auto and cross-spectra of the $95\,$GHz, $150\,$GHz, and $220\,$GHz maps with point sources detected at $>6.4$\,mJy at 150 GHz masked at all frequencies.
The uncertainties in the table are calculated from the diagonal elements of the covariance matrix, which includes noise and sample variance, but not beam or calibration errors. 
Due to the larger beam size at 95\,GHz, the 95x95\,GHz bandpowers are limited to $\ell < 8800$. 
}
\normalsize
\end{center}
\end{table*}

\section{Null tests}
\label{sec:jackknife}

We test the data for unknown systematic errors by running two null tests.
A null test consists of dividing the set of maps into two halves. 
The power spectrum of the difference between the maps of these two halves should be consistent with zero since all true astrophysical signals are canceled out. 
In practice, there can be slight amounts of residual power due to, for instance, small pointing differences.
We calculate the expectation for the tiny amount of remaining power by applying the same differencing process to simulations. 
Detecting a significant deviation from this expectation would signal the presence of a systematic error. 
Note that we only run new null tests for the combined \sptsz{} and \sptpol{} field; we do not rerun null tests for the fields that have been reused from G15. 
We look at the following data splits for systematic effects:

\begin{itemize}

\item Scan direction: We subtract left-going from right-going scans to test for potential systematics related to the telescope's motion. 
This test is also sensitive to incorrect detector time constants.

\item Time: We split the data based on when it was observed. 
We subtract data from the first half of the observations of a field from data from  the second half. 
Note that we split the data such that half 1 had the first half of the SPT-SZ observations plus the first half of the SPTpol observations, rather than all of the SPT-SZ observations plus some SPTpol observations. 
The null tests demonstrates the long-term temporal stability of the instruments. 
For instance, a slow drift in calibration would cause the test to fail.

\end{itemize}
We find one failure in the null tests. 
The first-second half null test at 150\,GHz shows excess power at $\ell < 2500$. 
While this excess is statistically significant (approximately $4\,\sigma$ in two bins), it is also extremely small, $<0.1$\% of the non-nulled power at these scales. 
Given the small amount of power, relative to either the bandpowers or the sample variance in these bins, we choose to proceed with the analysis.

\section{Bandpowers}
\label{sec:bandpowers}

\begin{figure*}[t]\centering
\includegraphics[trim=1.8cm 2.4cm 9cm 8cm,clip,width=0.9\textwidth]{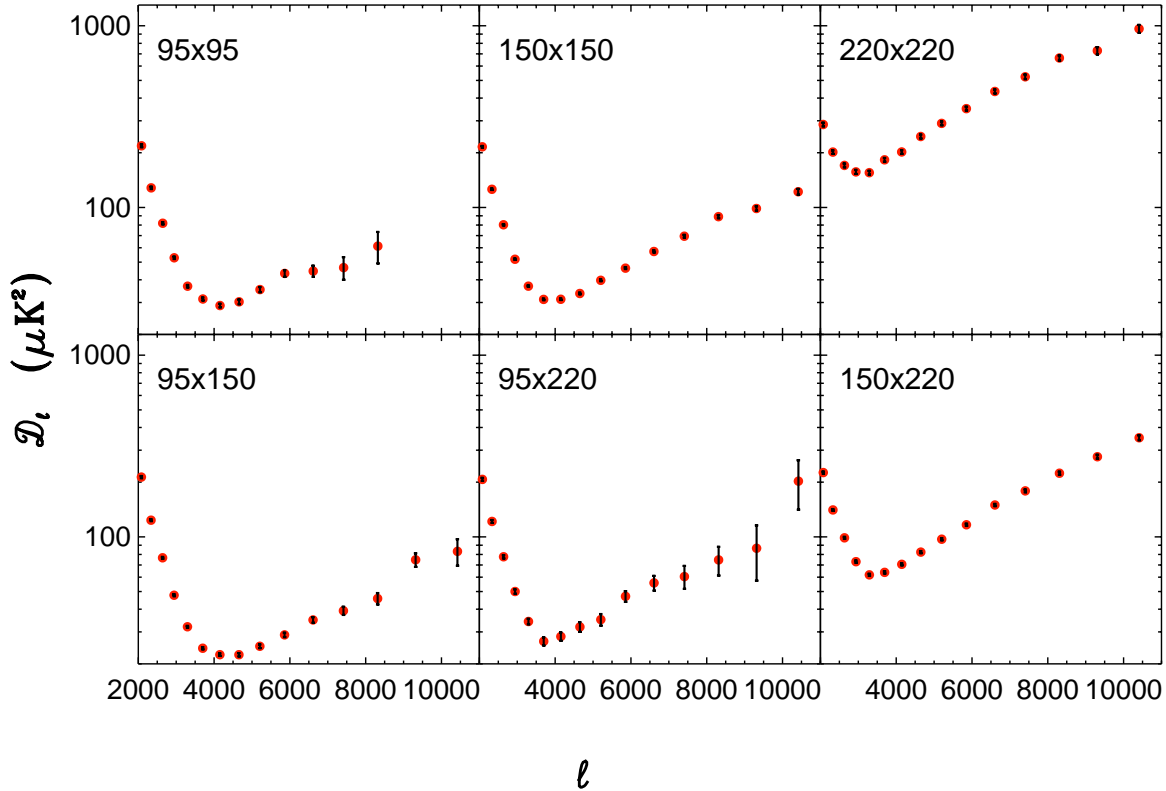}
\caption{  
The six auto- and cross-spectra measured with the 95, 150, and 220\,GHz SPT data. 
  }
  \label{fig:bandpower}
\end{figure*}

\begin{figure*}[t]\centering
\includegraphics[trim=1.8cm 1.8cm 9cm 8cm,clip,width=0.9\textwidth]{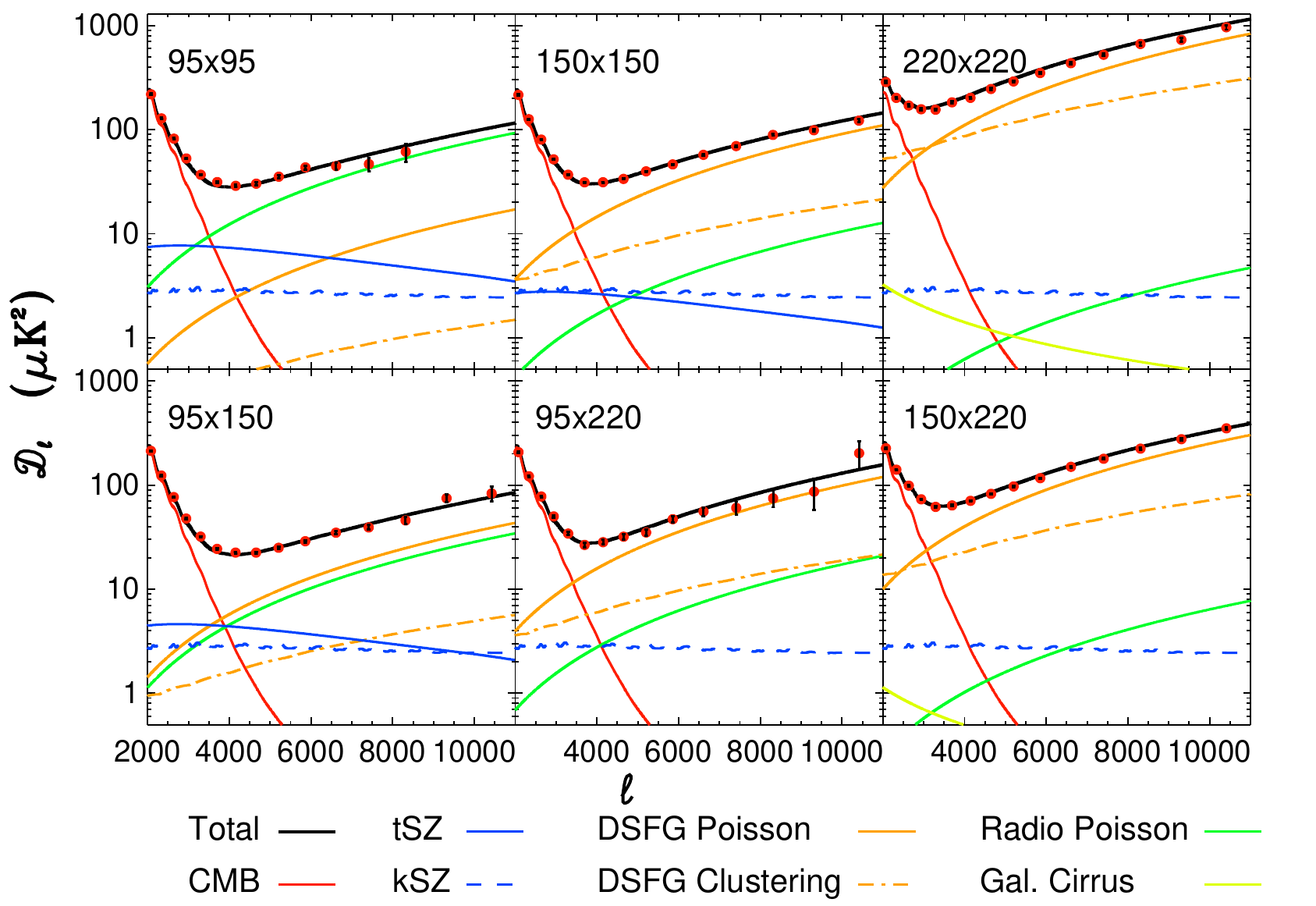}
\caption{  
The best-fit baseline model plotted against the SPT 95, 150, and 220\,GHz auto- and cross-spectra. 
We also show the relative power in each component of the model. 
  }
  \label{fig:bandpowerbestfit}
\end{figure*}

We apply the analysis of \S\ref{sec:bandpowerest} to the coadded \sptpol{} and \sptsz{} maps.  
Masking point sources above 6.4\,mJy at 150\,GHz leads to a final effective area of 464\,\sqdeg{} for the combined \sptpol{} plus \sptsz{} field.  
We combine the resulting bandpowers with those from the other 13 fields in G15 according to \S\ref{sec:fieldweighting}. 
As in G15, we measure the power spectra across the range of $2000 < \ell < 11,000$. 
Following G15, we restrict the 95$\times$95\,GHz bandpowers to $\ell <8800$ due to the larger beam size at 95\,GHz. 
The new bandpowers are listed in Table~\ref{tab:bandpowers} and plotted in Fig.~\ref{fig:bandpower}.
The bandpowers, covariance matrix, and window functions are available for download on the SPT\footnote{\label{footnote:spt}http://pole.uchicago.edu/public/data/reichardt20/} 
and LAMBDA\footnote{\label{footnote:lambda}http://lambda.gsfc.nasa.gov/product/spt/spt\_prod\_table.cfm} websites. 

The observed power is dominated by the primary CMB anisotropy on large angular scales ($\ell < 3500$). 
On smaller scales, extragalactic sources become important, DSFGs at 150 and 220\,GHz, and radio galaxies at 95\,GHz. 
We also see evidence for power from the kinematic and thermal SZ effects. 
We plot the best-fit model components against the bandpowers at the 6 frequency combinations in Fig.~\ref{fig:bandpowerbestfit}.


\begin{table}[ht!]
\begin{center}
\caption{\label{tab:deltachisq} Delta $\chi^2$ for model components}
\small
\begin{tabular}{cc|cc}
\hline\hline
\rule[-2mm]{0mm}{6mm}
term & dof & \delchisq\  \\
\hline
CMB (fixed) + Cirrus &- & (reference)\\
DSFG Poisson&2 & -77175.\\
Radio Poisson& 2 & -5135.\\
DSFG Clustering&3&  -985.\\ 
tSZ & 1 &  -269.\\  
kSZ $+$  tSZ-CIB Correlation & 2 & -8.4 \\
$\ell$-dependent tSZ-CIB & 0 & -0.5 \\
\hline
\hline
          Sloped tSZ-CIB corr. & 1 &  0.0\\  
                $T\in[8,50\,K]$ & 2 & +0.4 \\ 
                Scatter in spectral indices & 2 & +0.2 \\ 
      	 Power-law for cluster DSFG & 0 & +2.1\\ 
    Separate h- and p-kSZ & 1 & +0.8\\ 

\hline
\end{tabular}
\tablecomments{ Improvement to the best-fit $\chi^2$ as additional terms are added to the model. 
Terms above the double line are included in the baseline model, with each row showing the improvement in likelihood relative to the row above it. 
Note that adding either kSZ or a tSZ-CIB correlation separately leads to a marginal improvement in $\chi^2$ ($\Delta\chi^2\sim$ 1-2) but the improvement is more significant with both parameters included.  
For rows below the double line, the $\Delta\chi^2$ is shown relative to the baseline model rather than the row above it. 
None of these extensions significantly improve the fit quality. 
The row labeled ``Sloped tSZ-CIB corr." multiplies the Shang tSZ-CIB correlation template by term that varies linearly with $\ell$ around the pivot point of unity at $\ell=3000$.
The row labeled ``$T\in[8,50\,K]$" allows the temperature of the modified BB for the Poisson and clustered CIB terms to vary between 8 and 50\,K.  
The row labeled ``Scatter in spectral indices" adds two parameters, describing the population variance in spectral indices between CIB and radio galaxies respectively. 
The row labeled ``Power-law for cluster DSFG" replaces the one- and two-halo CIB templates by a power-law described by an amplitude and an exponent. 
While this conserves the total number of model parameters, the power law form is a worse fit to the data. 
Finally, in ``Separate h- and p-kSZ", we check if the data can distinguish between the (small) expected change in angular dependence between the homogeneous and patchy kSZ terms. 
Surprisingly, allowing two amplitude parameters, one for each kSZ template, results in a worse fit. 
The uncertainty on all of the quoted  $\delchisq$s is approximately 0.4.
} \normalsize
\end{center}
\end{table}

\section{Cosmological modeling}
\label{sec:model}

We fit the \sptpol{} + \sptsz{} bandpowers to a combination of the primary CMB anisotropy, thermal and kinematic SZ effects, radio galaxies, and DSFGs. 
The model is described in detail in the Appendix of G15; we only outline it here. 
The CMB is the most significant term on large angular scales in all bands. 
On smaller angular scales, the DSFGs contribute the most power at 150 and 220\,GHz, while radio galaxies are more significant at 95\,GHz.  
The SZ effects and correlations between the thermal SZ signal and CIB are also included. 
Finally, although the Galactic cirrus power in these fields and frequency bands is expected to be small, we include Galactic cirrus in our modeling, with an external prior on the amplitude and shape

We use the October 2019 version of {\textsc CosmoMC}\footnote{http://cosmologist.info/cosmomc} \citep{lewis02b} to calculate parameter constraints. 
We have added code to model the foregrounds and secondary anisotropies, which is based on the code used by G15. 
The source code and instructions to compile are available on the SPT website.$^{\ref{footnote:spt}}$

Unless otherwise noted, we fix the six \lcdm{} parameters to the best-fit values. 
The best-fit values are taken from a combined likelihood with the \planck{} 2018 TT, TE, and EE data, and the bandpowers of this work. 
We find that allowing  the \lcdm{} parameters to vary does not noticeably affect the recovered posteriors for the foreground and secondary anisotropy parameters, and the Monte Carlo Markov chain steps are much faster with the \lcdm{} parameters fixed.

Two terms in the modelling describe the kSZ and tSZ power spectra. 
We model the tSZ power as a free amplitude (defined by the power at $\ell=3000$ and 143 GHz) that scales the \citet{shaw10} tSZ model template. 
We assume the non-relativistic tSZ frequency scaling. 
In \S\ref{sec:ksztszbaseres}, we also check if the results depend on the template chosen. 
Similarly, we describe the kSZ power by an amplitude parameter (defined by the power at $\ell=3000$) that scales a template constructed by  setting the power of the CSF\footnote{Simulations that included cooling and star formation.} homogeneous kSZ template from \citet{shaw12} and patchy kSZ template from \citet[][hereafter Z12]{zahn12} to be equal at $\ell=3000$. 
Slightly differently than the tSZ case,  we test the data's sensitivity to the exact angular dependence of the kSZ power in \S\ref{sec:ksztszbaseres} by simultaneously fitting separate amplitudes for the homogeneous and patchy kSZ terms. 

We include two parameters to describe the radio Poisson power: the amplitude of the radio Poisson power at 150\,GHz and $\ell=3000$, and the spectral index $\alpha_{rg}$ for the radio  galaxies. 
Unlike in G15, we do not place a prior on the radio galaxy power as the 95\,GHz data constrains it well. 

In the baseline model, we include five parameters to describe the DSFGs that make up the CIB. 
Three of these parameters are amplitudes, respectively of the Poisson, one-halo clustering, and two-halo clustering power at $\ell=3000$ and 150\,GHz. 
As in G15, the one- and two-halo clustering templates are taken from the best-fit halo model in \citet{viero13a}. 
The other two parameters are the grey-body indices $\beta$ for the Poisson and clustering power respectively. 
We assume that there is no difference in the frequency scaling between the one- and two-halo clustering terms.

Finally, we include the expected anti-correlation between the CIB and tSZ power spectra. 
An anti-correlation is expected below the peak of the CMB black body because a dark matter over-density will be associated with an over-density of DSFGs (positive signal) and hot gas (negative tSZ signal). 
We take the angular dependence of the anti-correlation to be described by the form found by Z12, when looking at the \citet{shang12} CIB simulations. 
However, we allow the magnitude of this anti-correlation to float freely from -1 to 1, with the magnitude defined at $\ell=3000$. 

Table~\ref{tab:deltachisq} shows the improvement in the quality of the fits with the sequential
introduction of free parameters to the original $\Lambda$CDM primary CMB model. 
There are clear improvements as each parameter is added, up through the kSZ and tSZ-CIB correlation. 
Changing from a tSZ-CIB correlation that is constant in $\ell$ to the Z12 form marginally improves the $\chi^2$ by 0.5, without introducing any new parameters. 
Thus we include this shape in our baseline model. 
The other model variations we consider do not significantly improve the quality of the fits. 
Using a power-law for the CIB clustered power, as was done by G15 instead of the simulation-based 1- and 2-halo terms, is disfavored by the data, with an increase in $\chi^2$ of 2.1 for the same number of parameters.


\subsection{SPT effective frequencies}
\label{sec:efffreq}

While we refer to the three frequency bands as 95, 150, and 220\,GHz for convenience, the actual bandpasses are not simple delta functions. 
The bandpasses of both \sptpol{} and \sptsz{} were measured using a Fourier transform spectrometer (FTS). 
We estimate the calibration uncertainty on the FTS to be 0.3\,GHz, which should be coherent between the three bands. 
Although the uncertainty has negligible effect on the constraints, we  marginalize over the FTS calibration uncertainty in all parameter fits for completeness.

With the measured bandpasses in hand, we can calculate an effective band center for each of the potential signals: the thermal SZ effect, the CIB, and synchrotron sources. 
As we report and calibrate the bandpowers in CMB temperature units, the band center is irrelevant for sources with a CMB-like spectrum. 
We average the measured band centers for each year using that year's data relative weight to the final bandpowers. 
For an $\alpha = -0.5$ (radio-like) source spectrum, we find band centers of 93.5, 149.5, and $215.8\,$GHz. 
For an $\alpha = 3.5$ (dust-like) source spectrum, we find band centers of 96.9, 153.4, and $221.6\,$GHz. 
For a non-relativistic tSZ spectrum, we find band centers of 96.6, 152.3, and $220.1\,$GHz. 
The ratio of tSZ power in the 95\,GHz band to that in the $150\,$GHz band is 2.77; the 220\,GHz band has nearly zero tSZ power as it is well-matched to the the null in the tSZ spectrum near 217\,GHz. 
Note that we quote all tSZ power constraints at 143\,GHz for consistency with \planck, and all other model terms at 150\,GHz.

\section{Results}
\label{sec:results}

\subsection{Baseline model}
\label{sec:baseres}

We begin by presenting results for the baseline model discussed in Section \ref{sec:model}. 
This model includes the best-fit $\Lambda$CDM model plus ten parameters to describe foregrounds. 
Foreground parameters include the amplitudes of the tSZ power, kSZ power, radio galaxy Poisson power, CIB Poisson power, and CIB 1- and 2-halo clustered power; two parameters to describe the frequency dependence of the CIB terms; the tSZ-CIB correlation; and the spectral index of radio galaxies. 
The amplitude of  galactic cirrus is allowed to float within a strong prior.

We fit the 88 SPT bandpowers to the model described above. 
There are 78 degrees of freedom (dof), since the $\Lambda$CDM parameters are set to their best-fit values, essentially fixed by the \planck{} data, leaving the ten foreground model parameters. 
Jointly fitting the foreground terms and the $\Lambda$CDM parameters with \planck{} data  has little effect on derived foreground constraints. 
This baseline model fits the SPT data with a \chisq{} = 99.7, giving a PTE of 5.0\% for our 78 degrees of freedom, and provides the simplest interpretation of the data.


\subsubsection{CIB constraints}
\label{sec:cibbaseres}

The CIB is detected at very high significance, and is especially important at 220\,GHz. 
As highlighted in Table~\ref{tab:deltachisq}, adding the CIB terms to the model improve the fit quality by $\Delta \chi^2 \sim 77,000$. 
With the flux cut of $\sim$6.4\,mJy at 150\,GHz in this work, the Poisson CIB power is larger than the radio galaxy power by a factor of seven at 150\,GHz and a factor of 60 at 220\,GHz. 
The radio galaxy power is larger than the CIB power at 95\,GHz. 

At 150 GHz and $\ell=3000$, we find that the Poisson DSFG component has  power $D^{p}_{3000}=\DdgPoisson\, \uksq{}$ while the one- and two-halo DSFG clustering terms are $D^{1-halo}_{3000}=\DdgCluster \,\uksq{}$ and $D^{2-halo}_{3000}=\DdgClusterII \,\uksq{}$ respectively. 
At 220\,GHz, this scales to $D^{p, \,{\rm 220\,GHz}}_{3000}=\DdgPoissonTwoTwentyghz\, \uksq{}$, $D^{1-halo,\, {\rm 220\,GHz}}_{3000}=\DdgClusterTwoTwentyghz \,\uksq{}$, and $D^{2-halo,\, {\rm 220\,GHz}}_{3000}=\DdgClusterIITwoTwentyghz \,\uksq{}$.
The $\beta$ in the modified black body functional form of $\nu^\beta B_\nu(T)$ rises from \BetadgPoisson{} for the Poisson term to \BetadgCluster{} for the clustered terms. 
Cast as effective spectral indices from 150 to 220\,GHz, these values of $\beta$ translate to spectral indices of \alphaDGPOneFiveTwoTwo{} for the Poisson power and \alphaDGCOneFiveTwoTwo{} for the clustered power. 

The constraints from the baseline model are close to both theoretical expectations and previous work (e.g., \citealt{dunkley11}, G15).
When considering a similar foreground model (except for the angular dependence of the tSZ-CIB correlation), G15 found Poisson power levels of $7.59\pm 0.69\,\uksq{}$ and $63.4\pm 9.5\,\uksq{}$ at 150 and 220\,GHz.  
Note that since G15 reported powers at the effective frequency bandcenters instead of 150 and 220\,GHz,  to facilitate a comparison we have rescaled the reported numbers to 150 and 220\,GHz using the median spectral index in this work. 
These two sets of constraints agree very closely ($0.3\,\sigma$ or 3-5\%). 
It should be remembered that there is a large overlap between the underlying data, especially at 220\,GHz where only the relative weighting of the data has changed. 
For the clustered terms, the G15 numbers are $1.6\pm 0.9$\,\uksq{} and $1.7\pm 0.3$\,\uksq{} for the one- and two-halo terms. 
The agreement is still good: the one-halo term has increased by $0.7\,\sigma$ while the two-halo term dropped by a smaller amount. 
 The same trends continue at 220\,GHz: the one-halo term increases by $0.5\,\sigma$ while the two-halo term falls slightly. 
We note  that the recovered CIB clustering power in G15 was slightly lower than previous measurements. 
 Thus the shifts here move towards those earlier measurements. 
The inferred spectral indices of this work are also  within $1\,\sigma$ with the values in G15. 
 
 \subsubsection{Radio galaxy constraints}
\label{sec:rgbaseres}

Radio power is detected at high significance at 95 and 150\,GHz, with the addition of radio power to the model 
leading to a large improvement in the fit quality, namely  $\Delta\chi^2 = 5141$ for two parameters. 
As in G15, the data prefer slightly less radio galaxy power than predicted by the \citet{dezotti05} model for a 6.4\.mJy flux cut at 150\,GHz. 
The preferred radio power at $\ell=3000$ is $D^{r-150\times 150}_{3000}=\Drg \, \uksq{}$, about 25\% lower than the 1.28\,\uksq{} predicted. 
The population spectral index for the radio power is constrained to be \AlphaRG. 
This is $1\,\sigma$ lower than the median spectral index of -0.60 for synchrotron-classified sources reported by \citet{mocanu13}. 
This could be due to random chance, the 150\,GHz-only selection criteria for masking point sources in this work, or a tendency for the spectral index to flatten for the brightest 150\,GHz radio sources, as argued by \citet{mocanu13}.

\subsubsection{SZ power} 
\label{sec:ksztszbaseres}

\begin{figure}[thb]
\begin{center}
\resizebox{0.45\textwidth}{!}{
\includegraphics[width=1.0\textwidth,clip=true, trim = 1.15cm 11.35cm 8.16cm 4.09cm]{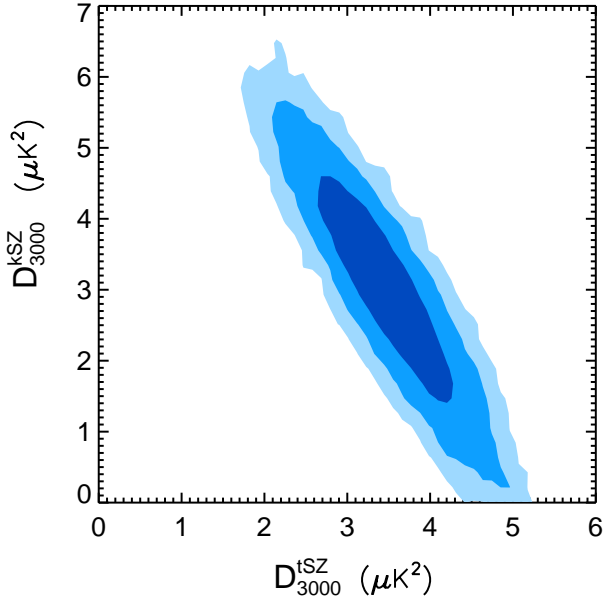}}
\end{center}
\caption
{2D posterior likelihood surface for the tSZ and kSZ power at 143 GHz at $\ell=3000$ in the baseline model including tSZ-CIB correlations. 
1, 2, and 3 $\sigma$ constraints are shown in shades of blue. 
The observed degeneracy is due to the correlation between the tSZ and CIB.
}
\label{fig:1paneltszksz}
\end{figure}

As shown in Figure~\ref{fig:1paneltszksz}, we detect both tSZ and kSZ power. 
We measure $\dtsz=\DtSZ\,\uksq{}$ and $\dksz =\DkSZ\,\uksq{}$ for the tSZ and kSZ power respectively at $\ell=3000$ and 143\,GHz. 

The tSZ (kSZ)  power is detected at approximately 7 (3)\,$\sigma$. 
While our fiducial results assume the Shaw tSZ template \citep{shaw10} and CSF+patchy kSZ template (Z12, \citealt{shaw12}), 
the current data offer little information about the specific shape of the SZ spectra. 
The recovered SZ power levels for four different tSZ templates and three kSZ templates are reported in Table \ref{tab:szconstraint} \-- no significant shifts are seen between the different templates considered. 
The tSZ power spectrum level is a probe of large-scale structure growth and the pressure profiles in galaxy clusters. 
The total kSZ power has contributions from the epoch of reionization and from the bulk flows of large-scale structure at later times; we discuss the implications of the kSZ measurement for reionization in \S\ref{sec:kszinterp}.

The joint analysis of the SPTpol and SPT-SZ surveys allows the first detection (at $3\,\sigma$) of kSZ power. 
The reported kSZ power in this work falls within the 95\% CL upper limits on kSZ power reported in previous works (e.g., \citet{dunkley13}, G15). 
G15 also report a central value when including a tSZ prior based on the bispectrum of $\dksz=2.9 \pm 1.3\,\uksq{}$, which agrees extremely well (although with 30\% larger uncertainties) with the value in this work.  
If we add the same bispectrum-based tSZ prior to the current results, we find  $\dksz=\DkSZBispec$\,\uksq{}, which translates to a  $3.1\,\sigma$ detection of kSZ power. 

The joint analysis also significantly reduces the measurement uncertainties on the tSZ power. 
This tSZ measurement is consistent  with ($<1\,\sigma$) earlier observations of the tSZ power scaled to 143\,GHz: $\dtsz=4.38_{-1.04}^{+0.83}$\,\uksq{} (G15); $\dtsz=4.20 \pm 1.37$\,\uksq{} (R12), and $\dtsz=3.9 \pm 1.7 \,\uksq{}$ \citep{dunkley13}. 
With the same bispectrum-based prior, the preferred tSZ power in this work is $\dtsz=\DtSZBispec{}$ \,\uksq{}.

\begin{table*}[ht!]
\begin{center}
\caption{\label{tab:szconstraint} SZ constraints}
\small
\begin{tabular}{cc|c|c|c}
\hline\hline
\rule[-2mm]{0mm}{6mm}
tSZ Template& kSZ Template  & $D^{\rm tSZ}_{3000} ~(\mu {\rm K}^2)$ & $D^{\rm kSZ}_{3000} ~(\mu {\rm K}^2)$ & $\xi$ \\
\hline
Shaw &CSF+patchy  & \DtSZ & \DkSZ & \tSZCib \\
Shaw &CSF  &  \DtSZKSZHomog & \DkSZKSZHomog & \tSZCibKSZHomog \\
Shaw &Patchy  & \DtSZKSZEoR & \DkSZKSZEoR & \tSZCibKSZEoR\\
Battaglia &CSF+patchy  & \DtSZTSZBattaglia & \DkSZTSZBattaglia & \tSZCibTSZBattaglia\\
Bhattacharya &CSF+patchy  & \DtSZTSZSuman & \DkSZTSZSuman & \tSZCibTSZSuman\\
Sehgal &CSF+patchy  & \DtSZTSZSehgal & \DkSZTSZSehgal & \tSZCibTSZSehgal\\
\hline\hline
Shaw w. Bispectrum &CSF+patchy  & \DtSZBispec & \DkSZBispec & \tSZCibBispec \\

\hline
\end{tabular}
\tablecomments{ 
Measured tSZ power, kSZ power and tSZ-CIB correlation at $\ell=3000$ (and 143\,GHz in the case of the tSZ) for different tSZ and kSZ models. 
 The results are robust to the assumed templates. 
The first two columns indicate which of three templates has been used for the tSZ and kSZ terms. 
In the case of the kSZ, the three templates are the CSF homogeneous kSZ template \citep{shaw12}, the patchy kSZ template (Z12), or the sum of both. 
 In the case of the tSZ, the three templates are taken from the Battaglia \citep{battaglia13a}, Shaw \citep{shaw10}, Bhattacharya \citep{bhattacharya12}, or Sehgal \citep{sehgal10} simulations. 
 The last row shows the results when a prior on the tSZ power based on the bispectrum measurement by \citet{crawford14} is added. 
}
\normalsize
\end{center}
\end{table*}

\subsubsection{tSZ-CIB correlation} 
\label{sec:tszcibbaseres}

\begin{figure*}[htb]\centering
\begin{center}
\resizebox{0.8\textwidth}{!}{
\includegraphics[width=1.0\textwidth,clip=true, trim =  2.06cm  11.97cm  9.32cm 10.22cm]{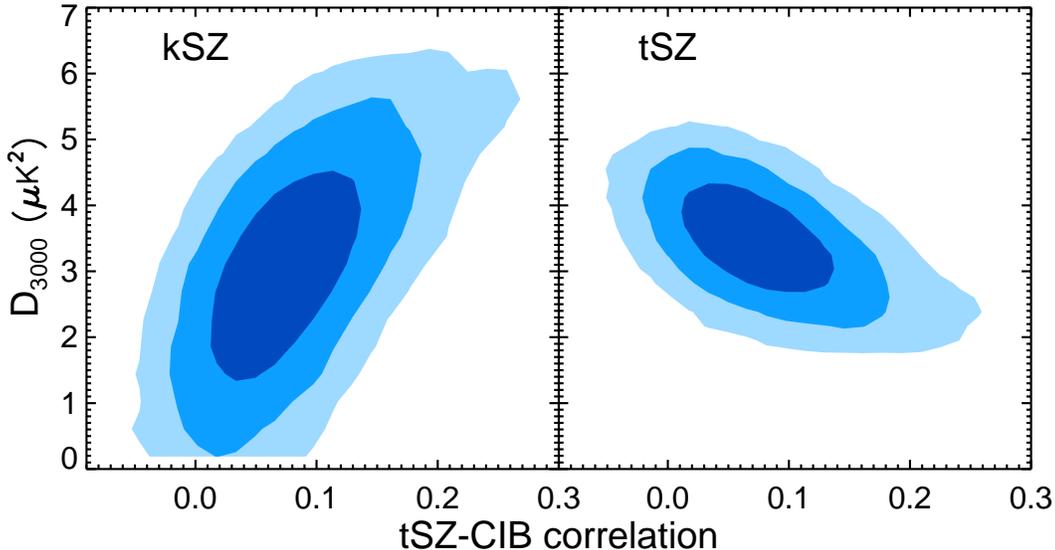}}
\end{center}
\caption
{ 
The  2D posterior likelihood of the tSZ-CIB correlation and kSZ (\textbf{\textit{Left panel}}) or 143\,GHz tSZ power (\textbf{\textit{Right panel}}). 
The filled contours show the 1, 2, and $3\,\sigma$ constraints. 
The data strongly prefer a positive tSZ-CIB correlation, consistent with DSFGs being over-dense in galaxy clusters. 
}
\label{fig:2paneltszcib}
\end{figure*}

We  parameterize the tSZ-CIB correlation with a single parameter $\xi$ that scales the Z12 template for the tSZ-CIB correlation as a function of $\ell$. 
An overdensity of dusty galaxies in galaxy clusters would result in a positive value of $\xi$. 
The tSZ-CIB correlation is partially degenerate with the tSZ and kSZ power, as illustrated in  
Figure \ref{fig:2paneltszcib}. 
Increasing the correlation, $\xi$,  slowly decreases the inferred tSZ power while quickly increasing the inferred kSZ power. 
We measure the tSZ-CIB correlation to be $\xi = \tSZCib$ at $\ell=3000$. 
The data prefer positive tSZ-CIB correlation, ruling out $\xi <0$ at the \PositiveTszCib{} CL. 
For easier comparison to past works, we also run a chain with $\xi$ that is constant in $\ell$. 
This does very little to the inferred SZ power levels; the preferred values shift by 0.2 and 0.3\,$\sigma$ for the tSZ and kSZ respectively. 
For a constant $\xi$, the bandpowers in this work favor \dksz= \DkSZFlatTszCib\,\uksq,  and \dtsz= \DtSZFlatTszCib\,\uksq. 
This is somewhat less (1\,$\sigma$) tSZ power than found by G15 in the equivalent case, and slightly more kSZ power (0.4\,$\sigma$). 
The $\xi$ constraint is $\xi = \tSZCibFlatTszCib$. 
This is well within $1\,\sigma$ of past SPT constraints, $\xi = 0.100^{+0.069}_{-0.055}$ (G15). 
It is also within the assumed prior range $[0,0.2]$ of \citet{dunkley13}.

\comment{\subsection{\tbd{}Constraints from the tSZ bispectrum} 
\label{sec:withbispec}

More information about the tSZ effect can be gleaned from higher-order moments of the map, such as the bispectrum. 
The bispectrum, and other higher-order statistics, are less affected by the kSZ-tSZ degeneracy because the kSZ effect is nearly Gaussian and thus contributes very little to the bispectrum. 
Adding bispectrum measurements can partially break that degeneracy, subject to modeling uncertainties in going from the bispectrum to power spectrum. 
Motivated by this potential, we investigate adding an external  tSZ power constraint derived from the 800 deg$^2$ SPT bispectrum measurement in C14. 
At 153.8\,GHz, C14 report $D_{3000}^\mathrm{tSZ-153.8\,GHz} = 2.96 \pm 0.64 \,\mu$K$^2$; this translates to $\dtsz = 3.81 \pm 0.82 \,\mu$K$^2$ at 143\,GHz.

As would be expected, the  bispectrum prior primarily impacts the SZ constraints, with a small impact on the amplitude of the tSZ-CIB correlation. 
There is no impact on the CIB constraints, with the largest shift being only 0.2\,$\sigma$. 
There is a small downward shift (0.3\,$\sigma$) in the median tSZ power from $\dtsz=4.38^{+0.83}_{-1.04} \ \,\uksq{}$ to $4.08^{+0.58}_{-0.67} \,\uksq{}$, and a 30\% reduction in the uncertainties. 
This in turn pushes the ML kSZ power higher and, when relaxing the positivity prior by running chains with negative values of kSZ power allowed, we find $\dksz{}= 2.87 \pm 1.25\,\uksq{}$. 
This translates to a 95\% CL upper limit of 4.9\,\uksq, and disfavors negative kSZ power at 98.1\% CL. 
When including the bispectrum measurement, the magnitude of the tSZ-CIB correlation is increased, going from $\xi = 0.100^{+0.069}_{-0.053}$ to $\xi = 0.113^{+0.057}_{-0.054}$.
This rejects anti-correlation ($\xi<0$) at 99.0\% CL. }

\section{kSZ interpretation} 
\label{sec:kszinterp}


The most significant improvement  in the current study compared to previous works is to the kSZ constraint, with the transition from upper limits to a $3\,\sigma$ detection of power. 
In this section, we look at what can be learned about the epoch of reionization from the kSZ measurement. 
We do this using the expression for the patchy kSZ power as a function of the timing and duration of reionization (among other cosmological parameters) presented by \cite{calabrese14}.

\subsection{Patchy kSZ power}

To interpret the measured kSZ power in light of the epoch of reionization, we must divide up the observed kSZ power between the homogeneous and patchy kSZ signals. 
As the current data can not separate the homogeneous kSZ and patchy kSZ power, we consider the inferred patchy kSZ power under three scenarios  for the homogeneous kSZ power. 
The estimate for the homogeneous kSZ power at $\ell=3000$ is taken from Eqn.~5 in \citet{calabrese14}, who in turn base it on the homogeneous kSZ simulations run by  \citet{shaw12}. 
 For the fiducial cosmology in this work, this estimate translates to $\dhksz = 1.65\,\uksq{}$.  
We also include high and low estimates of the homogeneous kSZ power, by rescaling the best guess by factors of 1.25 or 0.75 respectively. 
For comparison, \citet{shaw12} find that a different treatment of Helium reionization can scale the homogeneous kSZ signal by  $\sim$1.22 at $\ell=3000$. 

In all three cases, we take the shape of the homogeneous kSZ power from the CSF model in \citet{shaw12}. 
The angular dependence will change slightly for different models, for instance, different Helium ionization scenarios change the relative power between $\ell=3000$ and 10,000 by of order 3\%. 
However, the current data are insensitive to such small shape variations.

\comment{The lower bound is selected to be the CSF model from \citet{shaw12}, while the upper bound is chosen to the the NR model from that same work with Helium singly-ionized at z=6 and doubly ionized at z=3. 
As discussed by \citet{shaw12}, this prescription for Helium ionization increases the homogeneous kSZ signal by $\sim$1.22 at $\ell=3000$. 
While this factor changes slightly with angular multipole, we ignore these small shifts since the data are insensitive to the shape, and the variations ($\sim$3\% between $\ell=3000$ and 10,000) are much smaller than the current measurement uncertainties. 
}

With these assumptions about the homogeneous kSZ power in place, we find 95\% CL upper limits on the patchy kSZ power of $\dpksz<\pKSZCalabrese{}~ (\pKSZCalabreseHi/ \pKSZCalabreseLo)$\,\uksq{} for the best estimate of the homogeneous kSZ power (low/high homogeneous kSZ estimates). 
These limits on the patchy kSZ power are significantly better than the spectra-only limit of  $< 4.4\, \uksq$ reported by G15, and similar to what was achieved by the addition of the bispectrum prior in G15. 
If we add the same bispectrum prior to these chains while using the best-estimate of the homogeneous kSZ, the patchy kSZ upper limit  power  reduces by another 10\% to $\dpksz< \pKSZCalabreseBispec$\,\uksq. 
The 68\% confidence interval for the patchy kSZ power with the bispectrum prior is  $\dpksz= \pKSZCalabreseBispecInterval$\,\uksq. 


\subsection{Ionization history and the duration of reionization}

\begin{figure}[htb]\centering
\includegraphics[width=0.45\textwidth, clip=true, trim = 1.15cm 11.35cm 8.07cm 4.09cm]{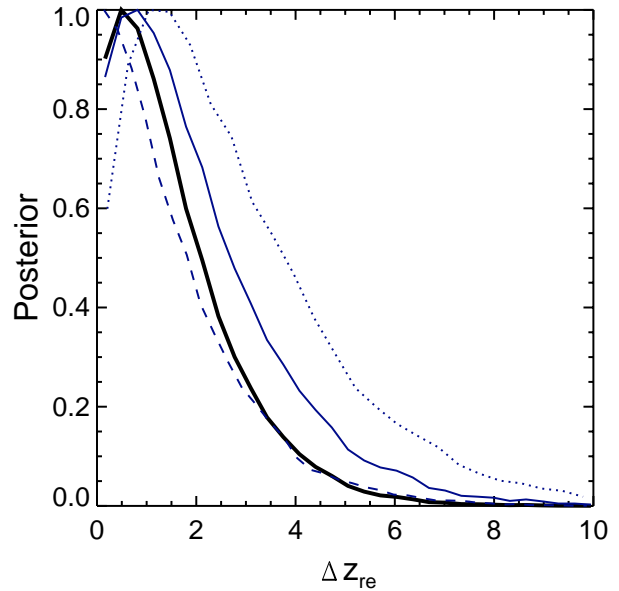}
  \caption{ 
1D likelihood curves for $\Delta z$ of reionization with the three assumptions about the homogeneous kSZ power used in this work. 
The solid blue line is for the expected amount of homogeneous kSZ power, while the dotted and dashed lines reflect the cases where the homogeneous kSZ power is scaled by $\times 0.75$ or 1.25 respectively. 
With the best estimate of the homogeneous kSZ power, the 95\% CL upper limit on the duration of reionization is $\Delta z < \DzEorCalabreseFlat$. 
Adding the tSZ bispectrum prior from \citet{crawford14} strengthens this limit to $\Delta z_{re} < \DzEorCalabreseFlatBispec$, as shown by the solid black line. 
 }
  \label{fig:dzplot}
\end{figure}

We can transform constraints on the inferred patchy kSZ power, under these assumptions for the homogeneous kSZ power, into constraints on the duration of EoR using the expression for  patchy kSZ power in Eqn. 6 of \citet{calabrese14}:
\begin{equation}
\dpksz = 2.03 \left[ \left(\frac{1+z_{re}}{11}\right) - 0.12\right] \left(\frac{\Delta z_{re}}{1.05}\right)^{0.51} \uksq,
\end{equation}
which is based on the models of \citet{battaglia13a}. 
Here $z_{re}$ is the redshift when the ionization fraction is 50\%, and $\Delta z_{re}$ is the duration of the epoch of reionization (EoR), defined as the period between 25\% and 75\% ionization fractions. 
We also have a choice of prior. 
For most of this work, results are quoted with a prior that is uniform in power or \dpksz{}. 
However in an upper limit regime given the relationship between $\Delta z_{re}$ and \dpksz{}, a flat prior on \dpksz{} preferentially favors $\Delta z_{re}$ near zero. 
We thus choose to report  $\Delta z_{re}$ constraints under a flat prior on  $\Delta z_{re}$ instead. 
With these assumptions about the homogeneous kSZ power and this prior on $\Delta z_{re}$  in place, we find 95\% CL upper limits on the duration of the EoR of $\Delta z_{re}<\DzEorCalabreseFlat ~(\DzEorCalabreseFlatLo\,/\, \DzEorCalabreseFlatHi)$ for the best estimate of the homogeneous  kSZ power (low/high homogeneous kSZ estimates). 
With the bispectrum-based prior on the tSZ power added, the limit becomes $\Delta z_{re}<\DzEorCalabreseFlatBispec$. 
The 68\% confidence interval is $\Delta z_{re} = \DzEorCalabreseFlatBispecInterval$. 
These limits agree with the recent picture from a variety of observations arguing that reionization happened fairly quickly. 
 Figure \ref{fig:dzplot} shows the likelihoods for $\Delta z_{re}$ of reionization.

 The limits quoted above on the duration of reionization are significantly better than the limits previously set by G15. 
 G15 found an upper limit on the duration of reionization of $\Delta z_{re}<5.4$, when including the bispectrum prior. 
 One should be cautious, however, in directly comparing the numbers  due to four model changes.
 First, G15 defined the duration from 20\% - 99\% ionization fraction, instead of the 25\% to 75\% in this work. 
 Second, G15 used a higher value for the optical depth from \wmap, which will drive the duration down by roughly a factor of 1.7 for a fixed level of patchy kSZ power. 
 Third, G15 used a uniform prior on the kSZ power instead of a uniform prior on $\Delta z_{re}$. 
 Finally, the fiducial homogeneous kSZ model in G15 predicted more power, approximately the high case in this work. 
 Given the degeneracy between the patch and homogeneous kSZ spectra, more homogeneous kSZ power translates to less patchy kSZ power and a shorter duration. 
 If we re-analyze the G15 bandpowers with the updated calibration, uniform prior in $\Delta z_{re}$, \planck{} optical depth, homogeous kSZ model and definition of duration in this work, we find the directly comparable 95\% CL upper limit with the bispectrum prior on the duration to be $\Delta z_{re}<\DzEorGeorgeCalabreseBispec$. 
 The directly comparable limit with the bispectrum information in this work of $\Delta z_{re}<\DzEorCalabreseFlatBispec$ is nearly a factor of two lower.

\section{Conclusions} 
\label{sec:conclusions}

We have presented improved measurements of the 95, 150 and 220\,GHz auto- and cross-spectra, created by combining data from the 2500\,\sqdeg{} SPT-SZ survey with the low-noise 500\,\sqdeg{} SPTpol survey. 
The combined data set substantially reduces the bandpower uncertainties over the last SPT release, especially in frequency combinations including 95\,GHz data.  
These bandpowers represent the most sensitive measurements of arcminute-scale anisotropy near the peak of the CMB blackbody spectrum. 

The signal at these frequencies and angular scales is composed of the primary CMB temperature anisotropy,  DSFGs, radio galaxies, and the kinematic and thermal SZ effects. 
We fit the data to a 10-parameter model for the DSFGs, radio galaxies and SZ effects (while fixing the primary CMB power spectrum to the best-fit values). 
For the first time, we find a $3\,\sigma$ detection of the kSZ power, with a level of \dksz = \DkSZ{}\,\uksq. 
The observed kSZ power can be deconstructed as the sum of the homogeneous and patchy kSZ terms, which are highly degenerate at current levels of sensitivity. 
However, using estimates of the homogeneous kSZ power from simulations, we calculate the residual patchy kSZ power and thus limits on the duration of reionization. 
We find a 95\% CL upper limit on the duration of reionization of $\Delta z_{re} < \DzEorCalabreseFlat$. 
Adding the tSZ bispectrum prior from \citet{crawford14} strengthens this limit to $\Delta z_{re} < \DzEorCalabreseFlatBispec$. 
The 68\% confidence interval is $\Delta z_{re} = \DzEorCalabreseFlatBispecInterval$.  
This supports the recent picture emerging from a number of sources that reionization happened at late times and fairly quickly.

The SPT is currently being used to conduct a five-year  survey of 1500\,\sqdeg{} with the SPT-3G camera. 
The final survey temperature noise levels are expected to be 3, 2, and 9 \ukarcmin{} for 95, 150, and 220\,GHz respectively \citep{bender18}, which will lead to substantially smaller uncertainties on the power spectrum in all six frequency combinations.   
Further in the future, the Simons Observatory and CMB-S4 will extend these measurements to larger sky areas, lower noise levels and more frequency bands \citep{simonsobs19,cmbs4-sb1}. 
Future CMB measurements should tightly constrain the reionization history of the Universe.


\acknowledgements{
The South Pole Telescope program is supported by the National Science Foundation through grants PLR-1248097 and OPP-1852617.
Partial support is also provided by the NSF Physics Frontier Center grant PHY-0114422 to the Kavli Institute of Cosmological Physics at the University of Chicago, the Kavli Foundation, and the Gordon and Betty Moore Foundation through grant GBMF\#947 to the University of Chicago.  
This work is also supported by the U.S. Department of Energy. 
SP acknowledges support from the Australian Research Council's Discovery Projects scheme (DP150103208). 
CR acknowledges support from Australian Research Council Centre of Excellence for All Sky Astrophysics in 3 Dimensions (ASTRO 3D), through project number CE170100013.
J.W.H. is supported by the National Science Foundation under Award No. AST-1402161. 
W.L.K.W is supported in part by the Kavli Institute for Cosmological Physics at the University of Chicago through grant NSF PHY-1125897 and an endowment from the Kavli Foundation and its founder Fred Kavli. 
B.B. is supported by the Fermi Research Alliance LLC under contract no. De-AC02-07CH11359 with the U.S. Department of Energy.  
The Cardiff authors acknowledge support from the UK Science and Technologies Facilities Council (STFC).
The CU Boulder group acknowledges support from NSF AST-0956135.  
The McGill authors acknowledge funding from the Natural Sciences and Engineering Research Council of Canada, Canadian Institute for Advanced Research, and the Fonds de Recherche du Qu\'ebec -- Nature et technologies.
The UCLA authors acknowledge support from NSF AST-1716965 and CSSI-1835865. 
AAS acknowledges support from NSF AST-1814719. 
Argonne National Lab, a U.S. Department of Energy Office of Science Laboratory, is operated by UChicago Argonne LLC under contract no. DE-AC02-06CH11357. 
We also acknowledge support from the Argonne  Center  for  Nanoscale  Materials. 
This research used resources of the National Energy Research Scientific Computing Center (NERSC), a U.S. Department of Energy Office of Science User Facility operated under Contract No. DE-AC02-05CH11231.
The data analysis pipeline also uses the scientific python stack \citep{hunter07, jones01, vanDerWalt11} and the HDF5 file format \citep{hdf5}.
}
\bibliographystyle{aasjournal}
\bibliography{../../BIBTEX/spt}

\end{document}

%% file: r19_baseline_parameters.tex
\newcommand{\DtSZ}{\ensuremath{ 3.42 \pm  0.54}}
\newcommand{\DtSZBispec}{\ensuremath{ 3.53 \pm  0.48}}
\newcommand{\DkSZ}{\ensuremath{  3.0 \pm   1.0}}
\newcommand{\DkSZBispec}{\ensuremath{  2.8 \pm   0.9}}
\newcommand{\DdgPoisson}{\ensuremath{ 7.24 \pm  0.63}}
\newcommand{\DdgPoissonTwoTwentyghz}{\ensuremath{ 61.4 \pm   9.0}}

\newcommand{\BetadgCluster}{\ensuremath{ 2.23 \pm  0.18}}
\newcommand{\BetadgPoisson}{\ensuremath{ 1.48 \pm  0.13}}
\newcommand{\DdgCluster}{\ensuremath{ 2.21 \pm  0.88}}
\newcommand{\DdgClusterTwoTwentyghz}{\ensuremath{ 32.4 \pm  11.2}}

\newcommand{\DdgClusterII}{\ensuremath{ 1.82 \pm  0.31}}
\newcommand{\DdgClusterIITwoTwentyghz}{\ensuremath{ 27.5 \pm   4.6}}
\newcommand{\alphaDGPOneFiveTwoTwo}{\ensuremath{ 3.29 \pm  0.13}}
\newcommand{\alphaDGCOneFiveTwoTwo}{\ensuremath{ 4.04 \pm  0.18}}
\newcommand{\AlphaRG}{\ensuremath{-0.76 \pm  0.15}}
\newcommand{\Drg}{\ensuremath{ 1.01 \pm  0.17}}
\newcommand{\tSZCib}{\ensuremath{0.076 \pm 0.040}}
\newcommand{\tSZCibBispec}{\ensuremath{0.069 \pm 0.036}}

%% file: r19_altsz_parameters.tex
\newcommand{\DtSZKSZEoR}{\ensuremath{ 3.45 \pm  0.56}}
\newcommand{\DkSZKSZEoR}{\ensuremath{  3.5 \pm   1.2}}
\newcommand{\tSZCibKSZEoR}{\ensuremath{0.086 \pm 0.050}}
\newcommand{\DtSZKSZHomog}{\ensuremath{ 3.39 \pm  0.58}}
\newcommand{\DkSZKSZHomog}{\ensuremath{  3.1 \pm   1.3}}
\newcommand{\tSZCibKSZHomog}{\ensuremath{0.077 \pm 0.047}}
\newcommand{\DtSZTSZSehgal}{\ensuremath{ 3.59 \pm  0.54}}
\newcommand{\DkSZTSZSehgal}{\ensuremath{  2.8 \pm   1.0}}
\newcommand{\tSZCibTSZSehgal}{\ensuremath{0.064 \pm 0.039}}
\newcommand{\DtSZTSZSuman}{\ensuremath{ 3.46 \pm  0.54}}
\newcommand{\DkSZTSZSuman}{\ensuremath{  3.0 \pm   1.0}}
\newcommand{\tSZCibTSZSuman}{\ensuremath{0.071 \pm 0.036}}
\newcommand{\DtSZTSZBattaglia}{\ensuremath{ 3.74 \pm  0.54}}
\newcommand{\DkSZTSZBattaglia}{\ensuremath{  2.4 \pm   1.0}}
\newcommand{\tSZCibTSZBattaglia}{\ensuremath{0.051 \pm 0.033}}
\newcommand{\DtSZFlatTszCib}{\ensuremath{ 3.30 \pm  0.64}}
\newcommand{\DkSZFlatTszCib}{\ensuremath{  3.5 \pm   1.2}}
\newcommand{\tSZCibFlatTszCib}{\ensuremath{0.078 \pm 0.049}}

%% file: r19_ksz_eor_parameters.tex
\newcommand{\pKSZCalabrese}{\ensuremath{  2.9 }}

\newcommand{\pKSZCalabreseBispec}{\ensuremath{  2.5 }}
\newcommand{\pKSZCalabreseBispecInterval}{\ensuremath{  1.1^{+  1.0}_{ -0.7}}}

\newcommand{\pKSZCalabreseHi}{\ensuremath{  3.4 }}

\newcommand{\pKSZCalabreseLo}{\ensuremath{  2.5 }}
\newcommand{\DzEorCalabreseFlat}{\ensuremath{  5.4 }}
\newcommand{\DzEorCalabreseFlatBispec}{\ensuremath{  4.1 }}
\newcommand{\DzEorCalabreseFlatBispecInterval}{\ensuremath{  1.1^{+  1.6}_{ -0.7}}}
\newcommand{\DzEorCalabreseFlatLo}{\ensuremath{  6.9 }}
\newcommand{\DzEorCalabreseFlatHi}{\ensuremath{  4.3 }}

\newcommand{\DzEorGeorgeCalabreseBispec}{\ensuremath{  8.5 }}

%% file: r19_probabilities.tex
\newcommand{\PositiveTszCib}{0.983} 

%% file: authors.tex
\shortauthors{C.~L.~Reichardt, S.~Patil, et al.}
\author[0000-0003-2226-9169]{C.~L.~Reichardt} \affiliation{School of Physics, University of Melbourne, Parkville, VIC 3010, Australia} \affiliation{ARC Centre of Excellence for All Sky Astrophysics in 3 Dimensions (ASTRO 3D)}
\author{S.~Patil} \affiliation{School of Physics, University of Melbourne, Parkville, VIC 3010, Australia}
\author{P.~A.~R.~Ade} \affiliation{Cardiff University, Cardiff CF10 3XQ, United Kingdom}
\author{A.~J.~Anderson} \affiliation{Fermi National Accelerator Laboratory, MS209, P.O. Box 500, Batavia, IL 60510}
\author{J.~E.~Austermann} \affiliation{NIST Quantum Devices Group, 325 Broadway Mailcode 817.03, Boulder, CO, USA 80305} \affiliation{Department of Physics, University of Colorado, Boulder, CO, USA 80309}
\author{J.~S.~Avva} \affiliation{Department of Physics, University of California, Berkeley, CA, USA 94720}
\author{E.~Baxter} \affiliation{Institute for Astronomy, University of Hawai’i, 2680 Woodlawn Drive, Honolulu, HI 96822, USA} \affiliation{Department of Physics and Astronomy, University of Pennsylvania, Philadelphia, PA 19104, USA}
\author{J.~A.~Beall} \affiliation{NIST Quantum Devices Group, 325 Broadway Mailcode 817.03, Boulder, CO, USA 80305}
\author{A.~N.~Bender} \affiliation{High Energy Physics Division, Argonne National Laboratory, 9700 S. Cass Avenue, Argonne, IL, USA 60439} \affiliation{Kavli Institute for Cosmological Physics, University of Chicago, 5640 South Ellis Avenue, Chicago, IL, USA 60637}
\author[0000-0002-5108-6823]{B.~A.~Benson} \affiliation{Fermi National Accelerator Laboratory, MS209, P.O. Box 500, Batavia, IL 60510} \affiliation{Kavli Institute for Cosmological Physics, University of Chicago, 5640 South Ellis Avenue, Chicago, IL, USA 60637} \affiliation{Department of Astronomy and Astrophysics, University of Chicago, 5640 South Ellis Avenue, Chicago, IL, USA 60637}
\author[0000-0003-4847-3483]{F.~Bianchini} \affiliation{School of Physics, University of Melbourne, Parkville, VIC 3010, Australia}
\author[0000-0001-7665-5079]{L.~E.~Bleem} \affiliation{High Energy Physics Division, Argonne National Laboratory, 9700 S. Cass Avenue, Argonne, IL, USA 60439} \affiliation{Kavli Institute for Cosmological Physics, University of Chicago, 5640 South Ellis Avenue, Chicago, IL, USA 60637}
\author{J.~E.~Carlstrom} \affiliation{Kavli Institute for Cosmological Physics, University of Chicago, 5640 South Ellis Avenue, Chicago, IL, USA 60637} \affiliation{Department of Physics, University of Chicago, 5640 South Ellis Avenue, Chicago, IL, USA 60637} \affiliation{High Energy Physics Division, Argonne National Laboratory, 9700 S. Cass Avenue, Argonne, IL, USA 60439} \affiliation{Department of Astronomy and Astrophysics, University of Chicago, 5640 South Ellis Avenue, Chicago, IL, USA 60637} \affiliation{Enrico Fermi Institute, University of Chicago, 5640 South Ellis Avenue, Chicago, IL, USA 60637}
\author{C.~L.~Chang} \affiliation{Kavli Institute for Cosmological Physics, University of Chicago, 5640 South Ellis Avenue, Chicago, IL, USA 60637} \affiliation{High Energy Physics Division, Argonne National Laboratory, 9700 S. Cass Avenue, Argonne, IL, USA 60439} \affiliation{Department of Astronomy and Astrophysics, University of Chicago, 5640 South Ellis Avenue, Chicago, IL, USA 60637}
\author{P.~Chaubal} \affiliation{School of Physics, University of Melbourne, Parkville, VIC 3010, Australia}
\author{H.~C.~Chiang} \affiliation{Department of Physics, McGill University, 3600 Rue University, Montreal, Quebec H3A 2T8, Canada} \affiliation{School of Mathematics, Statistics \& Computer Science, University of KwaZulu-Natal, Durban, South Africa}
\author{T.~L.~Chou} \affiliation{Department of Physics, University of Chicago, 5640 South Ellis Avenue, Chicago, IL, USA 60637}
\author{R.~Citron} \affiliation{University of Chicago, 5640 South Ellis Avenue, Chicago, IL, USA 60637}
\author{C.~Corbett~Moran} \affiliation{Jet Propulsion Laboratory, 4800 Oak Grove Drive, Pasadena, CA, USA 91109 USA}
\author[0000-0001-9000-5013]{T.~M.~Crawford} \affiliation{Kavli Institute for Cosmological Physics, University of Chicago, 5640 South Ellis Avenue, Chicago, IL, USA 60637} \affiliation{Department of Astronomy and Astrophysics, University of Chicago, 5640 South Ellis Avenue, Chicago, IL, USA 60637}
\author{A.~T.~Crites} \affiliation{Kavli Institute for Cosmological Physics, University of Chicago, 5640 South Ellis Avenue, Chicago, IL, USA 60637} \affiliation{Department of Astronomy and Astrophysics, University of Chicago, 5640 South Ellis Avenue, Chicago, IL, USA 60637} \affiliation{California Institute of Technology, MS 249-17, 1216 E. California Blvd., Pasadena, CA, USA 91125}
\author{T.~de~Haan} \affiliation{Department of Physics, University of California, Berkeley, CA, USA 94720} \affiliation{Physics Division, Lawrence Berkeley National Laboratory, Berkeley, CA, USA 94720}
\author{M.~A.~Dobbs} \affiliation{Department of Physics, McGill University, 3600 Rue University, Montreal, Quebec H3A 2T8, Canada} \affiliation{Canadian Institute for Advanced Research, CIFAR Program in Gravity and the Extreme Universe, Toronto, ON, M5G 1Z8, Canada}
\author{W.~Everett} \affiliation{Department of Astrophysical and Planetary Sciences, University of Colorado, Boulder, CO, USA 80309}
\author{J.~Gallicchio} \affiliation{Kavli Institute for Cosmological Physics, University of Chicago, 5640 South Ellis Avenue, Chicago, IL, USA 60637} \affiliation{Harvey Mudd College, 301 Platt Blvd., Claremont, CA, USA 91711}
\author[0000-0001-7874-0445]{E.~M.~George} \affiliation{European Southern Observatory, Karl-Schwarzschild-Str. 2, 85748 Garching bei M\"{u}nchen, Germany} \affiliation{Department of Physics, University of California, Berkeley, CA, USA 94720}
\author{A.~Gilbert} \affiliation{Department of Physics, McGill University, 3600 Rue University, Montreal, Quebec H3A 2T8, Canada}
\author{N.~Gupta} \affiliation{School of Physics, University of Melbourne, Parkville, VIC 3010, Australia}
\author{N.~W.~Halverson} \affiliation{Department of Astrophysical and Planetary Sciences, University of Colorado, Boulder, CO, USA 80309} \affiliation{Department of Physics, University of Colorado, Boulder, CO, USA 80309}
\author{N.~Harrington} \affiliation{Department of Physics, University of California, Berkeley, CA, USA 94720}
\author{J.~W.~Henning} \affiliation{High Energy Physics Division, Argonne National Laboratory, 9700 S. Cass Avenue, Argonne, IL, USA 60439} \affiliation{Kavli Institute for Cosmological Physics, University of Chicago, 5640 South Ellis Avenue, Chicago, IL, USA 60637}
\author{G.~C.~Hilton} \affiliation{NIST Quantum Devices Group, 325 Broadway Mailcode 817.03, Boulder, CO, USA 80305}
\author[0000-0002-0463-6394]{G.~P.~Holder} \affiliation{Astronomy Department, University of Illinois at Urbana-Champaign, 1002 W. Green Street, Urbana, IL 61801, USA} \affiliation{Department of Physics, University of Illinois Urbana-Champaign, 1110 W. Green Street, Urbana, IL 61801, USA} \affiliation{Canadian Institute for Advanced Research, CIFAR Program in Gravity and the Extreme Universe, Toronto, ON, M5G 1Z8, Canada}
\author{W.~L.~Holzapfel} \affiliation{Department of Physics, University of California, Berkeley, CA, USA 94720}
\author{J.~D.~Hrubes} \affiliation{University of Chicago, 5640 South Ellis Avenue, Chicago, IL, USA 60637}
\author{N.~Huang} \affiliation{Department of Physics, University of California, Berkeley, CA, USA 94720}
\author{J.~Hubmayr} \affiliation{NIST Quantum Devices Group, 325 Broadway Mailcode 817.03, Boulder, CO, USA 80305}
\author{K.~D.~Irwin} \affiliation{SLAC National Accelerator Laboratory, 2575 Sand Hill Road, Menlo Park, CA 94025} \affiliation{Dept. of Physics, Stanford University, 382 Via Pueblo Mall, Stanford, CA 94305}
\author{L.~Knox} \affiliation{Department of Physics, University of California, One Shields Avenue, Davis, CA, USA 95616}
\author{A.~T.~Lee} \affiliation{Department of Physics, University of California, Berkeley, CA, USA 94720} \affiliation{Physics Division, Lawrence Berkeley National Laboratory, Berkeley, CA, USA 94720}
\author{D.~Li} \affiliation{NIST Quantum Devices Group, 325 Broadway Mailcode 817.03, Boulder, CO, USA 80305} \affiliation{SLAC National Accelerator Laboratory, 2575 Sand Hill Road, Menlo Park, CA 94025}
\author{A.~Lowitz} \affiliation{Department of Astronomy and Astrophysics, University of Chicago, 5640 South Ellis Avenue, Chicago, IL, USA 60637}
\author{D.~Luong-Van} \affiliation{University of Chicago, 5640 South Ellis Avenue, Chicago, IL, USA 60637}
\author{J.~J.~McMahon} \affiliation{Department of Physics, University of Michigan, 450 Church Street, Ann  Arbor, MI, USA 48109}
\author{J.~Mehl} \affiliation{Kavli Institute for Cosmological Physics, University of Chicago, 5640 South Ellis Avenue, Chicago, IL, USA 60637} \affiliation{Department of Astronomy and Astrophysics, University of Chicago, 5640 South Ellis Avenue, Chicago, IL, USA 60637}
\author{S.~S.~Meyer} \affiliation{Kavli Institute for Cosmological Physics, University of Chicago, 5640 South Ellis Avenue, Chicago, IL, USA 60637} \affiliation{Department of Physics, University of Chicago, 5640 South Ellis Avenue, Chicago, IL, USA 60637} \affiliation{Department of Astronomy and Astrophysics, University of Chicago, 5640 South Ellis Avenue, Chicago, IL, USA 60637} \affiliation{Enrico Fermi Institute, University of Chicago, 5640 South Ellis Avenue, Chicago, IL, USA 60637}
\author{M.~Millea} \affiliation{Department of Physics, University of California, One Shields Avenue, Davis, CA, USA 95616}
\author{L.~M.~Mocanu} \affiliation{Kavli Institute for Cosmological Physics, University of Chicago, 5640 South Ellis Avenue, Chicago, IL, USA 60637} \affiliation{Department of Astronomy and Astrophysics, University of Chicago, 5640 South Ellis Avenue, Chicago, IL, USA 60637}
\author{J.~J.~Mohr} \affiliation{Max-Planck-Institut f"ur extraterrestrische Physik,Giessenbachstr.\ 85748 Garching, Germany} \affiliation{Excellence Cluster Universe, Boltzmannstr.\ 2, 85748 Garching, Germany}
\author{J.~Montgomery} \affiliation{Department of Physics, McGill University, 3600 Rue University, Montreal, Quebec H3A 2T8, Canada}
\author{A.~Nadolski} \affiliation{Astronomy Department, University of Illinois at Urbana-Champaign, 1002 W. Green Street, Urbana, IL 61801, USA} \affiliation{Department of Physics, University of Illinois Urbana-Champaign, 1110 W. Green Street, Urbana, IL 61801, USA}
\author{T.~Natoli} \affiliation{Department of Astronomy and Astrophysics, University of Chicago, 5640 South Ellis Avenue, Chicago, IL, USA 60637} \affiliation{Kavli Institute for Cosmological Physics, University of Chicago, 5640 South Ellis Avenue, Chicago, IL, USA 60637} \affiliation{Dunlap Institute for Astronomy \& Astrophysics, University of Toronto, 50 St George St, Toronto, ON, M5S 3H4, Canada}
\author{J.~P.~Nibarger} \affiliation{NIST Quantum Devices Group, 325 Broadway Mailcode 817.03, Boulder, CO, USA 80305}
\author{G.~Noble} \affiliation{Department of Physics, McGill University, 3600 Rue University, Montreal, Quebec H3A 2T8, Canada}
\author{V.~Novosad} \affiliation{Materials Sciences Division, Argonne National Laboratory, 9700 S. Cass Avenue, Argonne, IL, USA 60439}
\author{Y.~Omori} \affiliation{Dept. of Physics, Stanford University, 382 Via Pueblo Mall, Stanford, CA 94305} \affiliation{Kavli Institute for Particle Astrophysics and Cosmology, Stanford University, 452 Lomita Mall, Stanford, CA 94305}
\author{S.~Padin} \affiliation{Kavli Institute for Cosmological Physics, University of Chicago, 5640 South Ellis Avenue, Chicago, IL, USA 60637} \affiliation{Department of Astronomy and Astrophysics, University of Chicago, 5640 South Ellis Avenue, Chicago, IL, USA 60637} \affiliation{California Institute of Technology, MS 249-17, 1216 E. California Blvd., Pasadena, CA, USA 91125}
\author{C.~Pryke} \affiliation{School of Physics and Astronomy, University of Minnesota, 116 Church Street S.E. Minneapolis, MN, USA 55455}
\author{J.~E.~Ruhl} \affiliation{Physics Department, Center for Education and Research in Cosmology and Astrophysics, Case Western Reserve University, Cleveland, OH, USA 44106}
\author{B.~R.~Saliwanchik} \affiliation{Physics Department, Center for Education and Research in Cosmology and Astrophysics, Case Western Reserve University, Cleveland, OH, USA 44106} \affiliation{Department of Physics, Yale University, P.O. Box 208120, New Haven, CT 06520-8120}
\author{J.T.~Sayre} \affiliation{Department of Astrophysical and Planetary Sciences, University of Colorado, Boulder, CO, USA 80309} \affiliation{Department of Physics, University of Colorado, Boulder, CO, USA 80309} \affiliation{Physics Department, Center for Education and Research in Cosmology and Astrophysics, Case Western Reserve University, Cleveland, OH, USA 44106}
\author{K.~K.~Schaffer} \affiliation{Kavli Institute for Cosmological Physics, University of Chicago, 5640 South Ellis Avenue, Chicago, IL, USA 60637} \affiliation{Enrico Fermi Institute, University of Chicago, 5640 South Ellis Avenue, Chicago, IL, USA 60637} \affiliation{Liberal Arts Department, School of the Art Institute of Chicago, 112 S Michigan Ave, Chicago, IL, USA 60603}
\author{E.~Shirokoff} \affiliation{Kavli Institute for Cosmological Physics, University of Chicago, 5640 South Ellis Avenue, Chicago, IL, USA 60637}
\author{C.~Sievers} \affiliation{University of Chicago, 5640 South Ellis Avenue, Chicago, IL, USA 60637}
\author{G.~Smecher} \affiliation{Department of Physics, McGill University, 3600 Rue University, Montreal, Quebec H3A 2T8, Canada} \affiliation{Three-Speed Logic, Inc., Victoria, B.C., V8S 3Z5, Canada}
\author{H.~G.~Spieler} \affiliation{Physics Division, Lawrence Berkeley National Laboratory, Berkeley, CA, USA 94720}
\author{Z.~Staniszewski} \affiliation{Physics Department, Center for Education and Research in Cosmology and Astrophysics, Case Western Reserve University, Cleveland, OH, USA 44106}
\author{A.~A.~Stark} \affiliation{Center for Astrophysics $|$ Harvard \& Smithsonian, 60 Garden Street, Cambridge, MA, USA 02138}
\author{C.~Tucker} \affiliation{Cardiff University, Cardiff CF10 3XQ, United Kingdom}
\author{K.~Vanderlinde} \affiliation{Dunlap Institute for Astronomy \& Astrophysics, University of Toronto, 50 St George St, Toronto, ON, M5S 3H4, Canada} \affiliation{Department of Astronomy \& Astrophysics, University of Toronto, 50 St George St, Toronto, ON, M5S 3H4, Canada}
\author{T.~Veach} \affiliation{Department of Astronomy, University of Maryland College Park, MD, USA 20742}
\author{J.~D.~Vieira} \affiliation{Astronomy Department, University of Illinois at Urbana-Champaign, 1002 W. Green Street, Urbana, IL 61801, USA} \affiliation{Department of Physics, University of Illinois Urbana-Champaign, 1110 W. Green Street, Urbana, IL 61801, USA}
\author{G.~Wang} \affiliation{High Energy Physics Division, Argonne National Laboratory, 9700 S. Cass Avenue, Argonne, IL, USA 60439}
\author[0000-0002-3157-0407]{N.~Whitehorn} \affiliation{Department of Physics and Astronomy, University of California, Los Angeles, CA, USA 90095}
\author{R.~Williamson} \affiliation{University of Chicago, 5640 South Ellis Avenue, Chicago, IL, USA 60637} \affiliation{Department of Astronomy and Astrophysics, University of Chicago, 5640 South Ellis Avenue, Chicago, IL, USA 60637}
\author[0000-0001-5411-6920]{W.~L.~K.~Wu} \affiliation{Kavli Institute for Cosmological Physics, University of Chicago, 5640 South Ellis Avenue, Chicago, IL, USA 60637}
\author{V.~Yefremenko} \affiliation{High Energy Physics Division, Argonne National Laboratory, 9700 S. Cass Avenue, Argonne, IL, USA 60439}